\documentclass{article} 
\usepackage{iclr2024_conference,times}

\usepackage{amsmath,amsfonts,bm}

\def\eqref#1{equation~\ref{#1}}

\def\1{\bm{1}}

\DeclareMathAlphabet{\mathsfit}{\encodingdefault}{\sfdefault}{m}{sl}
\SetMathAlphabet{\mathsfit}{bold}{\encodingdefault}{\sfdefault}{bx}{n}

\usepackage{hyperref}
\usepackage{url}

\usepackage[utf8]{inputenc} 
\usepackage[T1]{fontenc}   
\usepackage{hyperref}       
\usepackage{url}           
\usepackage{booktabs}       
\usepackage{amsfonts}       
\usepackage{nicefrac}       
\usepackage{microtype}     
\usepackage{xcolor}

\usepackage{url}
\usepackage{amsmath}
\usepackage{amssymb}
\usepackage{mathtools}
\usepackage{amsthm}
\usepackage{multirow,tabularx}
\usepackage{longtable}
\usepackage{booktabs,longtable}
\usepackage{hhline}
\usepackage{microtype}
\usepackage{graphicx}
\usepackage{booktabs}
\usepackage{algorithmic}
\usepackage{algorithm}
\usepackage{pifont}
\usepackage{enumitem}
\usepackage{tcolorbox}
\usepackage{wrapfig}
\usepackage{amsmath}
\usepackage{bm}
\usepackage{fontawesome}
\usepackage{cancel} %
\usepackage[normalem]{ulem} %
\usepackage{makecell}
\usepackage{cleveref}

\title{
Where the Body Keeps the Beat: Structured Motion Conditioning and Music Dynamics Supervision for Dance-to-Music Generation
}

\author{Changchang Sun$^{1}$ ~~~~Lu Cheng$^{2}$ ~~~~\textbf{Yan Yan$^1$} \\
  $^1$University of Illinois Chicago
  ~~~~~ $^2$Penn State University	
  \\
}

\iclrfinalcopy
\date{}
\begin{document}

\maketitle
\fancyhead{}
\fancyhead[C]{\normalfont\ttfamily Dyna2Music}
\renewcommand{\headrulewidth}{0.4pt}
\thispagestyle{plain} \renewcommand{\iclrruler}[1]{}

\begin{abstract}
  \label{sec: abstract}
{Dance-to-music (D2M) generation aims to synthesize musically plausible soundtracks whose temporal structure aligns with a given dance performance.
Representative D2M methods do not explicitly distinguish motion cues across body parts and frequency bands, while standard flow matching lacks a dedicated objective for supervising local music-latent dynamics.
To address these issues, we propose \textbf{Dyna2Music}, a latent flow-matching framework that combines structured motion conditioning with explicit supervision of music-latent dynamics.
An empirical analysis on AIST++ quantifies how spatial partitioning and frequency separation affect raw music-to-kinematic beat alignment and motion-reference density, informing the conditioning design.
Accordingly, Dyna2Music decomposes joint velocities into slow and fast components and hierarchically fuses the resulting part-wise motion energy with pretrained joint features to condition music generation.
To complement this representation, we introduce latent dynamics consistency (LDC), an auxiliary objective that matches adjacent-frame change magnitudes between a single-step clean-latent estimate and the paired reference.
LDC makes local music-latent variation an explicit training target without adding trainable parameters or inference computation.
Dyna2Music supports variable-length music generation, and experiments on AIST++ and TikTok demonstrate improved rhythmic alignment and audio quality over representative D2M baselines.}
\end{abstract}

\section{Introduction}
\label{sec: introduction}

\newcommand{\introrev}[1]{#1}

For dance videos shared on platforms such as \href{https://www.tiktok.com/}{TikTok} and \href{https://www.youtube.com/}{YouTube}, the choice of background music shapes both the expressive quality of the performance and the viewing experience.
This application motivates research into dance-to-music (D2M) generation~\citep{gan2020foley,di2021video,zhu2022quantized,zhu2022discrete,yu2023long,han2024dance2midi,li2024dance,zhang2024dance2music,liang2024dancecomposer,sun2025enhancing,ji2025comprehensive}.
Unlike music generation for general video content, which addresses audiovisual compatibility through visual-semantic conditioning~\citep{tian2025vidmuse} and 
rhythmic controls derived from visual motion speed and saliency~\citep{di2021video}, 
D2M places particular emphasis on synthesizing musically plausible soundtracks whose temporal structure aligns with a given dance performance. Beyond matching the overall style or tempo, the soundtrack should respond to changes in movement, including sustained body motion, brief limb articulations, and rhythmic accents. {Addressing these requirements involves both representing dance dynamics and learning the temporal structure of the accompanying music.}

Existing diffusion-based D2M approaches typically condition music generation on visual features and pose-derived motion cues extracted from dance videos.
Within this paradigm, LORIS~\citep{yu2023long} derives a single visual rhythm sequence from aggregated joint motion, while PN-Diffusion~\citep{sun2025enhancing} uses a spatiotemporal graph convolutional network to encode skeletal sequences into a motion embedding.
However, the resulting {conditioning representations} do not explicitly separate cues by body part or temporal frequency.
\textit{Spatially}, different body parts may provide complementary cues about musical beat timing.
Aggregating motion across parts can obscure localized events, such as a brief wrist accent occurring alongside a larger torso movement, making these cues less distinguishable in the conditioning representation.
\introrev{To examine this issue, we analyze $8{,}819$ paired training clips from AIST++~\citep{li2021aichoreographer}, a dataset of dance videos paired with music that is widely used as a benchmark for D2M generation~\citep{zhu2022discrete,yu2023long,sun2025enhancing}.}
\begin{wrapfigure}{r}{0.62\textwidth}
\centering
\vspace*{-4mm}
\includegraphics[width=1.0\linewidth]{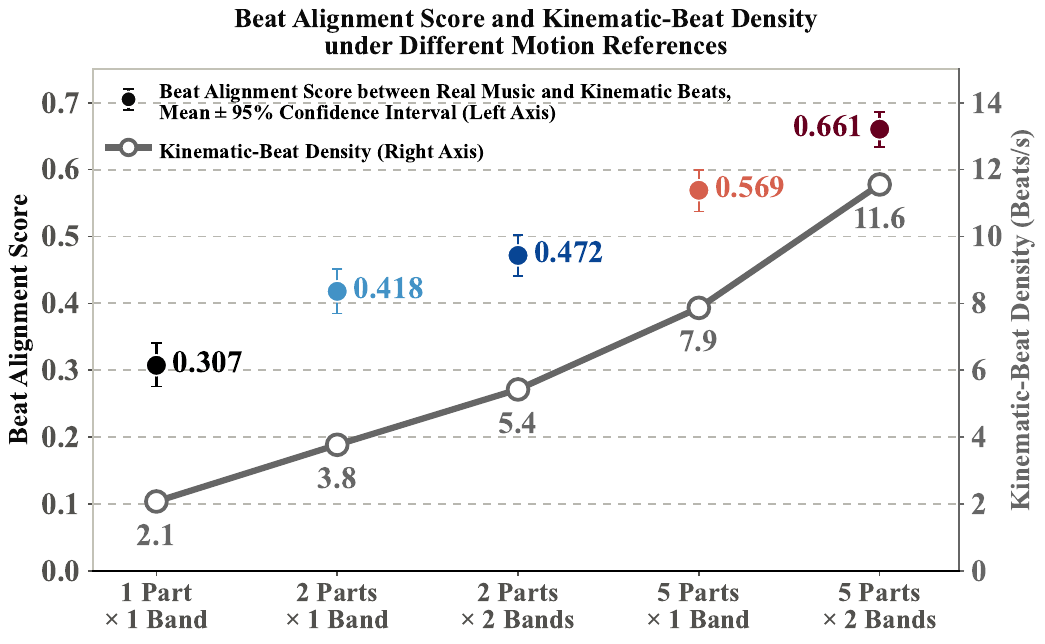}
\vspace*{-8mm}
\caption{
\introrev{
Beat alignment score (BAS) and kinematic-beat density for different motion references on paired dance-music clips from the AIST++ training set. Filled markers and error bars (left axis) indicate mean music-to-kinematic BAS and $95\%$ confidence intervals, respectively; the grey line (right axis) shows kinematic-beat density (beats/s). The spatial partitions comprise the whole body (one part), upper and lower body (two parts), or the torso, left and right arms, and left and right legs (five parts). One band denotes motion without frequency separation; two bands denote separate slow and fast components. Finer spatial partitions and slow-fast separation yield denser motion references, helping to explain the increase in raw BAS.
}
}
\label{fig: bas}
\vspace*{-5mm}
\end{wrapfigure}
{Figure~\ref{fig: bas} compares whole-body, two-part, and five-part motion references, each formed by taking the union of motion-derived beat candidates (kinematic beats) from its constituent signals.
We use the music-to-kinematic beat alignment score (BAS)~\citep{siyao2022bailando} to measure each music beat's temporal proximity to its nearest kinematic beat.
Without frequency separation, raw BAS increases from $0.307$ to $0.418$ and $0.569$ as the spatial partition becomes finer.}
\textit{Temporally}, slowly varying dynamics and rapid local fluctuations may convey complementary information about musical tempo, beat phase, and metrical subdivisions. Although existing methods capture temporal dependencies, they do not explicitly represent {low-frequency (slow) and high-frequency (fast)} motion components as distinct conditioning signals.
{Separating slow and fast components further increases raw BAS to $0.472$ and $0.661$ for the two- and five-part references, respectively; the five-part reference thus scores higher in both settings.}
\introrev{These raw scores also depend on motion-reference density, as illustrated in Figure~\ref{fig: bas}.}
We therefore investigate how {part-specific slow and fast motion signals} relate to musical rhythm, using the findings to inform the conditioning design.

Beyond representing rhythmic cues in dance, D2M generation also depends on how temporal variation in the accompanying music is learned.
For generators operating in audio latent space, the magnitudes of changes between adjacent latent frames characterize local temporal variation in the audio representation.
The standard flow-matching objective~\citep{lipman2023flow} regresses the generative velocity field without a dedicated term for comparing these dynamics between the predicted clean latent and the paired reference.
We therefore augment velocity regression with an auxiliary objective that matches these change magnitudes at corresponding music-latent frames.
This provides explicit supervision of local audio-latent dynamics, complementing part-specific slow and fast motion conditioning.

\begin{figure}[!t]
\centering
\centerline{\includegraphics[width=1.0\columnwidth]{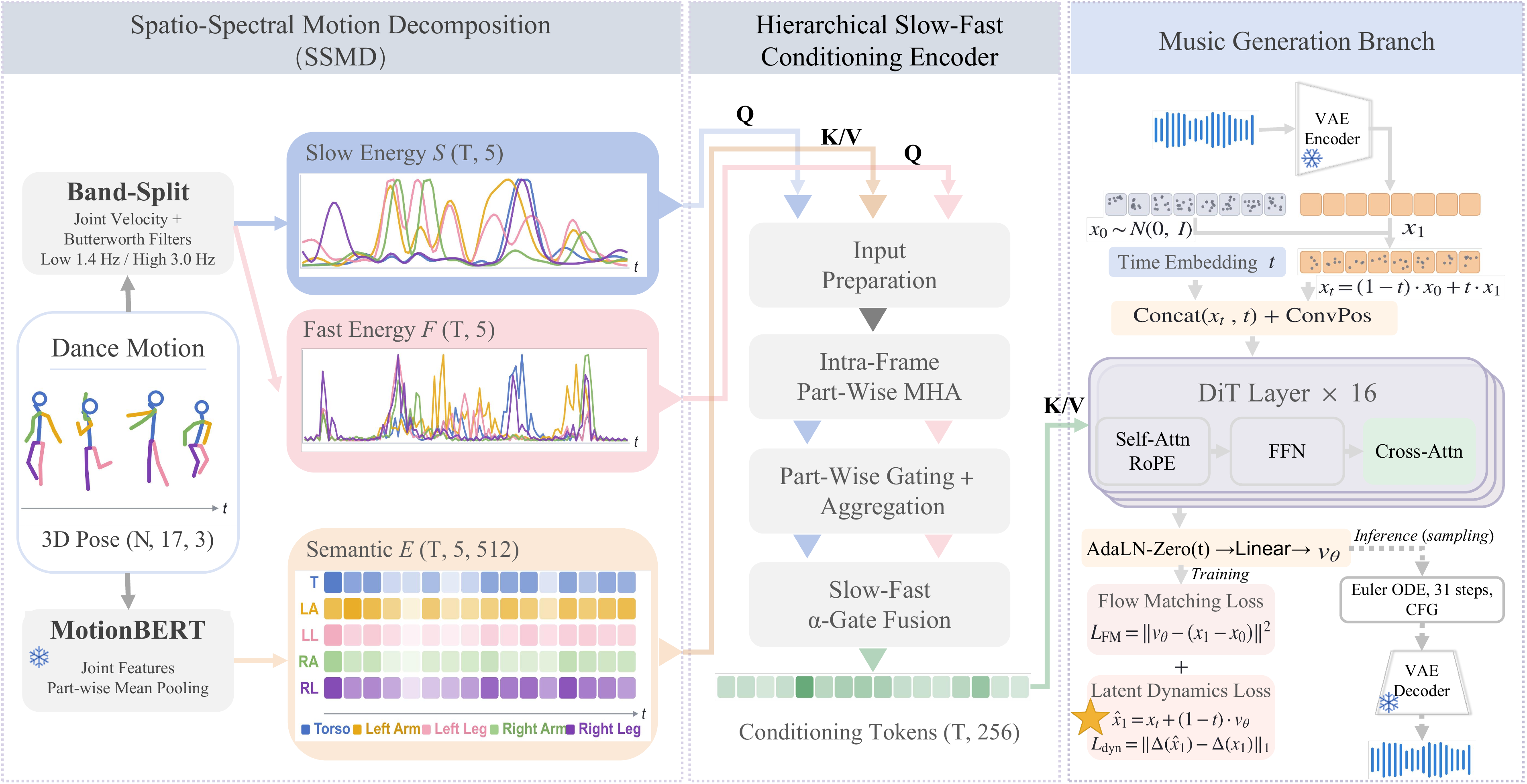}}
     \vspace*{-4mm}
     \caption{
\textbf{Overview of Dyna2Music.}
\textbf{Left:} SSMD extracts part-wise slow/fast motion energy
and complementary features from frozen MotionBERT.
\textbf{Middle:} The HSF-Encoder fuses these signals into
conditioning tokens through attention and adaptive gating.
\textbf{Right:} The tokens guide a latent flow-matching
transformer trained with velocity regression and latent
dynamics consistency (LDC), which supervises adjacent-frame
change magnitudes. Generated latents are decoded into music
by a frozen waveform VAE.
}
\label{fig: framework}
 \vspace*{-6mm}
\end{figure}

To this end, we introduce \textbf{Dyna2Music}, a latent
flow-matching framework for D2M generation that combines
structured motion conditioning with explicit supervision of
music-latent dynamics (Figure~\ref{fig: framework}).
To preserve motion cues across body parts and temporal scales,
the \textbf{Spatio-Spectral Motion Decomposition (SSMD)}
module separates joint velocities into low- and high-frequency
components and computes their energy for each of five body parts.
These energy signals are complemented by pretrained MotionBERT
features~\citep{zhu2023motionbert}, pooled within each part to
provide information about joint configurations and coordination
that energy alone does not capture.
The \textbf{Hierarchical Slow-Fast Conditioning Encoder
(HSF-Encoder)} then integrates the energy signals and motion
features through attention across body parts and adaptive gating,
producing conditioning tokens that combine part-specific context
with slow and fast motion dynamics.
The resulting conditioning tokens guide a transformer
adapted from DiffRhythm~\citep{ning2025diffrhythm} through cross-attention
to generate continuous music latents, which are decoded into
music by a frozen waveform VAE.
{To complement structured motion conditioning, we introduce \textbf{latent dynamics consistency (LDC)} as an auxiliary objective for supervising music-latent dynamics.
During training, we derive a single-step estimate of the clean music latent from the predicted velocity field.
LDC matches the magnitudes of adjacent-frame changes in this estimate to those in the paired reference latent at corresponding music time steps.
This explicitly supervises the temporal profile of local latent variation, complementing the standard flow-matching objective.
The loss is computed entirely in latent space, requires no waveform decoding during training, and introduces neither additional trainable parameters nor additional computation at inference.}
We summarize our contributions as follows.
\begin{itemize}[leftmargin=*, itemsep=2pt,
               topsep=0pt, partopsep=0pt, parsep=0pt]
\item {\textbf{An empirical study of motion references.} We quantify how body-part partitioning and slow-fast separation affect raw music-to-kinematic beat alignment and kinematic-beat density on AIST++, informing the design and evaluation of motion conditioning.}
\item {\textbf{Structured motion conditioning for D2M generation.} We propose Dyna2Music, a latent flow-matching framework that combines explicit motion decomposition across body parts and frequency bands with hierarchical fusion of motion energy and pretrained joint features.}
\item {\textbf{Explicit supervision of music-latent dynamics.} We introduce LDC, an auxiliary objective that matches adjacent-frame change magnitudes between a single-step clean-latent estimate and the paired reference. This complements velocity regression with explicit supervision of local temporal variation, without additional trainable parameters or inference computation.}
\item {\textbf{Variable-length generation and empirical validation.} Dyna2Music supports music generation for dance sequences of varying durations. Experiments on AIST++ and TikTok demonstrate improved rhythmic alignment and audio quality over representative D2M baselines.}
\end{itemize}

\section{Related Work}
\label{sec: related work}

\noindent
{\textbf{General Video-to-Music Generation.}
It targets background music generation for diverse content, including films, vlogs, and documentaries, emphasizing semantic compatibility, temporal alignment, and controllability.
An early approach, CMT~\citep{di2021video}, controls symbolic music generation by mapping visual timing, motion speed, and motion saliency to beats, note density, and note strength.
For richer visual context, VidMuse~\citep{tian2025vidmuse} integrates local clip information and global video context through long-short-term modeling, while GVMGen~\citep{zuo2025gvmgen} selects visual features with learnable music queries and conditions generation through temporal cross-attention.
For explicit control, MTCV2M~\citep{wu2025controllable} combines fine-grained feature selection, progressive temporal alignment, and dynamic fusion of time-varying beat, melody, intensity, and emotion conditions.
Complementary approaches use pretrained multimodal and language models to link video understanding with music generation.
M$^2$UGen~\citep{liu2023m2ugen} integrates modality-specific encoders with LLaMA~2 to condition MusicGen or AudioLDM~2 on inputs including videos.
Similarly, SONIQUE~\citep{zhang2025sonique} converts video descriptions into musical tags for conditional diffusion, enabling soundtracking with unpaired video and music data.
More broadly, AudioX~\citep{tian2026audiox} conditions a diffusion transformer on fused text, video, and audio representations, supporting video-to-music generation in a unified multimodal framework.
These studies explore learned visual representations, explicit musical controls, and language interfaces for general video soundtracking.}

\noindent
{\textbf{Autoregressive Dance-to-Music Generation.}}
{Autoregressive D2M methods sequentially predict symbolic events or discrete audio tokens conditioned on motion and preceding outputs.
Early approaches include Dance2Music~\citep{aggarwal2021dance2music}, which generates notes online from pose and note histories, and RhythmicNet~\citep{su2021does}, which converts motion-derived rhythm into percussion and melodic accompaniment.
For multi-instrument generation, Dance2MIDI~\citep{han2024dance2midi} combines autoregressive drum generation with BERT-based track completion, while DanceComposer~\citep{liang2024dancecomposer} progressively introduces rhythm and style constraints.
UniMuMo~\citep{yang2025unimumo} extends this paradigm to unified text, music, and motion generation through a shared music-motion codebook and mutually conditioned autoregressive streams, supporting motion-to-music generation.
Across representations, decoding retains sequential dependencies between output time steps.}

\noindent
{\textbf{Non-autoregressive Dance-to-Music Generation.}}
{Non-autoregressive D2M methods use adversarial synthesis or iterative refinement without left-to-right token decoding.
D2M-GAN~\citep{zhu2022quantized} conditions vector-quantized adversarial generation on video and motion, whereas CDCD~\citep{zhu2022discrete} introduces contrastive learning into discrete diffusion.
In continuous latent space, \citet{tan2023motion} incorporate motion and genre conditions, while Dance2Music-Diffusion~\citep{zhang2024dance2music} uses a dedicated motion encoder.
For richer conditioning, LORIS~\citep{yu2023long} integrates visual, rhythmic, and genre context for long-term synthesis, while PN-Diffusion~\citep{sun2025enhancing} uses positive and negative rhythmic signals from forward and reversed dance sequences.
To adapt pretrained diffusion models, encoder-based textual inversion~\citep{li2024dance} maps dance rhythm and genre to text embeddings, whereas \citet{kimura2026dance} combine beat-guided contrastive pretraining with a ControlNet adapter for a frozen AudioLDM backbone.
Within this family, Dyna2Music combines part-specific slow-fast motion conditioning with explicit LDC supervision of local music-latent dynamics in a latent flow-matching framework.}

\section{Methodology}
\label{sec: method}
\subsection{Problem Formulation}
\label{subsec: problem_formulation}

Dyna2Music conditions music generation on the dancer's motion,
represented as a sequence of estimated 3D joint positions
$\mathbf{J}$ spanning $N$ video frames.
The objective is to synthesize a musically plausible accompaniment
whose temporal structure aligns with the observed dance dynamics.
We formulate this task as conditional generation in music latent space.
For each dance-music pair, a frozen pretrained
waveform VAE~\citep{ning2025diffrhythm} encodes the music waveform
$\mathbf{y}$ as
\begin{equation}
\mathbf{x}_1=\mathcal{E}(\mathbf{y})
\in\mathbb{R}^{T\times d_a},
\label{eq: d2m}
\end{equation}
where $T$ denotes the number of music-latent frames and $d_a$
is the latent dimensionality.
We seek to learn the conditional distribution
$p_{\theta}(\mathbf{x}_1\mid\mathbf{J})$.
{At inference, generated music latents are converted into waveforms by the frozen VAE decoder $\mathcal{D}$.}
Throughout this section, $i$ indexes video frames,
$n$ indexes music-latent frames, and $t\in[0,1]$ denotes flow time.

\subsection{Spatio-Spectral Motion Decomposition}
\label{subsec: ssmd}
SSMD constructs motion-energy signals and joint features for the five-part partition examined in Figure~\ref{fig: bas}: the torso, left and right arms, and left and right legs.
Let $\mathcal{P}$ denote these parts and $\mathcal{J}_p$ the joint index set of part $p$ within the 17-joint skeleton.

\noindent\textbf{Part-specific motion energy.}
{We compute joint velocities from consecutive poses and apply temporal low-pass and high-pass filters with cutoff frequencies of $1.4$\,Hz and $3.0$\,Hz to obtain slow and fast velocity sequences $\mathbf{v}^{s}$ and $\mathbf{v}^{f}$, respectively.}
For band $b\in\{s,f\}$, we average the squared speed of the joints within each part:
\begin{equation}
K^{b}_{i,p}=\frac{1}{|\mathcal{J}_p|}
\sum_{j\in\mathcal{J}_p}\|\mathbf{v}^{b}_{i,j}\|_2^2.
\label{eq: motion_energy}
\end{equation}
{Frequency separation is applied to joint velocities before energy computation and aggregation within each body part.
Here, slow and fast refer to the temporal frequency of velocity variations, rather than the magnitude of joint velocity.}

\noindent\textbf{Joint features and temporal alignment.}
A frozen MotionBERT backbone~\citep{zhu2023motionbert} processes the full-body pose sequence to extract 512-dimensional joint features, which we average within the same parts to provide complementary motion context.
We resample the energy signals and pooled features to the $T$ music-latent frames, obtaining $\mathbf{S},\mathbf{F}\in\mathbb{R}^{T\times5}$ and $\mathbf{E}\in\mathbb{R}^{T\times5\times512}$ as inputs to the HSF-Encoder.
The joint partition, filtering, and preprocessing details are specified in Appendix~\ref{subsec: app_motion}.

\subsection{Hierarchical Slow-Fast Conditioning Encoder}
\label{subsec: hsf}
{As shown in Figure~\ref{fig: framework}, the HSF-Encoder maps $\mathbf{S}$, $\mathbf{F}$, and $\mathbf{E}$ to a conditioning sequence $\mathbf{C}\in\mathbb{R}^{T\times d}$, where $d=256$.}
{At each music-latent time step, the encoder retains distinct tokens for each part and band during context integration, uses energy-dependent gates to aggregate parts within each band, and then fuses the slow and fast representations.
}

\noindent\textbf{{Input Preparation.}}
{Energy sequences are standardized over valid time steps independently for each part and band, while joint features undergo layer normalization.
Both are then projected to $d$ dimensions and augmented with temporal and part embeddings; additional band embeddings distinguish the slow and fast energy streams.
The resulting representations are denoted by $\widetilde{\mathbf{B}}^s$, $\widetilde{\mathbf{B}}^f$, and $\widetilde{\mathbf{E}}$.}
The embedding definitions and normalization details are provided in Appendix~\ref{subsec: app_hsf}; affine biases are omitted from the notation for brevity.

\noindent\textbf{{Intra-Frame Part-Wise MHA.}}
{At each music-latent frame $n$, multi-head attention (MHA) integrates joint-feature context across the five body parts using band-specific energy embeddings as queries:}
\begin{equation}
\mathbf{H}^{b}_n=
\mathrm{MHA}_{b}\!\left(\widetilde{\mathbf{B}}^b_n,
\widetilde{\mathbf{E}}_n,\widetilde{\mathbf{E}}_n\right)
\in\mathbb{R}^{5\times d}.
\label{eq: slow_and_fast_2}
\end{equation}
{The three arguments denote queries, keys, and values, respectively.
Query projections are band-specific, whereas key, value, and output projections are shared across bands.}
{This design allows each part's band-specific energy representation to retrieve motion context from all body parts before aggregation.}

\noindent\textbf{{Part-Wise Gating + Aggregation.}}
{For each band, a gating network maps the energy embeddings to scores that are normalized across parts.
The resulting weights aggregate the attention outputs into a frame-level representation:}
\begin{equation}
 g^b_{n,p}=\mathrm{softmax}_{p}\!\left(
 \mathrm{MLP}^{\mathrm{gate}}_b(\widetilde{\mathbf{B}}^b_{n,p})\right),
 \qquad
 \mathbf{h}^{b}_n=\sum_{p\in\mathcal{P}}g^b_{n,p}\mathbf{H}^{b}_{n,p}.
\label{eq: slow_and_fast_3}
\end{equation}

\noindent\textbf{{Slow-Fast $\alpha$-Gate Fusion.}}
{A learned scalar gate $\alpha_n$ controls the relative contributions of the slow and fast representations to the mixture $\mathbf{m}_n$.
The final projection combines this mixture with their normalized concatenation $\mathbf{u}_n$:}
\begin{equation}
\begin{aligned}
\mathbf{u}_n&=\mathrm{LN}([\mathbf{h}^{s}_n;\mathbf{h}^{f}_n]),
&\alpha_n&=\sigma(\mathbf{w}_{\alpha}^{\mathsf T}\mathbf{u}_n),\\
\mathbf{m}_n&=\alpha_n\mathbf{h}^{s}_n+(1-\alpha_n)\mathbf{h}^{f}_n,
&\mathbf{C}_n&=\mathrm{LN}\!\left(\mathbf{W}_{F}[\mathbf{m}_n;\mathbf{u}_n]
+\mathbf{e}^{\mathrm{time}}_n\right).
\end{aligned}
\label{eq: hsf_fusion}
\end{equation}
{Retaining the concatenated branch representations alongside their gated mixture gives the final projection access to both branches separately.}
{Here, $\mathrm{LN}$ denotes layer normalization, $\sigma$ is the sigmoid function, $[\cdot;\cdot]$ denotes concatenation, and $\mathbf{e}^{\mathrm{time}}_n$ is the temporal positional embedding.
The resulting sequence $\mathbf{C}$ provides the motion conditions for the music generator.}

\subsection{Latent Flow Matching with Dynamics Supervision}
\label{subsec: flow_generator}
\label{subsec: ldc}
Conditioned on the HSF sequence $\mathbf{C}$, a transformer adapted from DiffRhythm~\citep{ning2025diffrhythm} predicts the music-latent velocity $\mathbf{v}_{\theta}(\mathbf{x}_t,t,\mathbf{C})$ (Figure~\ref{fig: framework}).
In each block, music representations query $\mathbf{C}$ through cross-attention; architecture details are provided in Appendix~\ref{subsec: app_backbone}.

\noindent\textbf{Conditional flow matching.}
Following conditional flow matching~\citep{lipman2023flow}, we interpolate Gaussian noise $\mathbf{x}_0$ with the paired music latent $\mathbf{x}_1$ to define the path and target velocity:
\begin{equation}
\mathbf{x}_t=(1-t)\mathbf{x}_0+t\mathbf{x}_1,
\qquad
\mathbf{v}^{\star}=\mathbf{x}_1-\mathbf{x}_0.
\label{eq: flow_path}
\end{equation}
The flow-matching loss $\mathcal{L}_{\mathrm{FM}}$ averages squared errors between predicted and target velocities over valid frames and channels.
Flow-time sampling and the masked loss reduction are specified in Appendix~\ref{subsec: app_sampling}.

\noindent\textbf{Latent dynamics consistency.}
LDC complements velocity regression by explicitly supervising local temporal variation in music latents.
It uses a single-step estimate of the clean music latent:
\begin{equation}
\widehat{\mathbf{x}}_1=
\mathbf{x}_t+(1-t)\mathbf{v}_{\theta}(\mathbf{x}_t,t,\mathbf{C}).
\label{eq: clean_estimate}
\end{equation}
For a latent sequence $\mathbf{z}$, the adjacent-frame change magnitude is the root mean square across channels:
\begin{equation}
\Delta_n(\mathbf{z})=
\sqrt{\frac{1}{d_a}\sum_{k=1}^{d_a}
(z_{n,k}-z_{n-1,k})^2+\varepsilon},
\qquad n\geq2,\quad\varepsilon=10^{-6}.
\label{eq: latent_change}
\end{equation}
Let $q$ index minibatch samples and $\mathcal{I}_{\Delta}$ contain pairs $(q,n)$ for which both $n-1$ and $n$ are valid.
The auxiliary loss and full objective are
\begin{equation}
\begin{aligned}
\mathcal{L}_{\mathrm{LDC}}
&=\frac{\displaystyle\sum_{(q,n)\in\mathcal{I}_{\Delta}}
\left|\Delta_n(\widehat{\mathbf{x}}_{1,q})-
\Delta_n(\mathbf{x}_{1,q})\right|}
{\max(1,|\mathcal{I}_{\Delta}|)},\\
\mathcal{L}&=\mathcal{L}_{\mathrm{FM}}+\lambda_{\mathrm{LDC}}\mathcal{L}_{\mathrm{LDC}}.
\end{aligned}
\label{loss}
\end{equation}
Here, $\lambda_{\mathrm{LDC}}\geq0$ controls the auxiliary supervision, with zero recovering flow matching alone.
LDC matches local change magnitudes at corresponding music frames; it does not directly constrain change directions or beat locations.
It uses the same velocity prediction and adds neither trainable parameters nor inference computation.
The HSF-Encoder and music transformer are trained jointly, while MotionBERT and the waveform VAE remain frozen.

\noindent\textbf{Inference and variable-length generation.}
The dance duration and VAE temporal resolution determine the latent length $T$ within the configured sequence-length limit.
We integrate the conditional velocity field from Gaussian noise using Euler updates and decode the resulting latents with the frozen VAE.
Sampling details are given in Appendix~\ref{subsec: app_sampling}.

\section{Experiments}
\label{sec: experiment}

\subsection{Datasets}
\label{subsec: datasets}
\begingroup

We evaluate Dyna2Music on two complementary D2M benchmarks,
AIST++~\citep{li2021aichoreographer,tsuchida2019aist} and
TikTok~\citep{zhu2022quantized}, covering dance performances
captured in controlled studio environments and in-the-wild
social-media videos, respectively.
Together, they allow us to assess rhythmic alignment and
audio quality across different recording conditions.
(i) \textbf{AIST++}.
This dataset contains dance performances recorded in a controlled studio environment.
Our split comprises $8{,}819$ training and $180$ test dance video-music pairs.
(ii) \textbf{TikTok}.
This dataset contains in-the-wild dance videos collected from social media, with diverse backgrounds and recording conditions.
Our split comprises $391$ training and $53$ test dance video-music pairs.

Training-length eligibility and sequence caps are detailed in Appendix~\ref{subsec: app_backbone}.
\par
\endgroup

\begin{table*}[!tbp]
\centering
\vspace{-4mm}
\caption{Onset-level synchrony (\%) on AIST++ and TikTok.
BCS/BHS denote precision/recall; CSD/HSD are their across-clip standard deviations.
Best values are bold.}
\label{tab:current-rhythm-comparison}
\small
\setlength{\tabcolsep}{3.5pt}
\renewcommand{\arraystretch}{1.05}
\scalebox{0.7}{
\begin{tabular}{@{}c|c|*{5}{c}|*{5}{c}@{}}
\toprule
\multirow{2}{*}{Category} & \multirow{2}{*}{Method}
& \multicolumn{5}{c|}{AIST++} & \multicolumn{5}{c}{TikTok} \\
\cmidrule(lr){3-7}\cmidrule(l){8-12}
& & BCS$\uparrow$ & CSD$\downarrow$ & BHS$\uparrow$ & HSD$\downarrow$ & F1$\uparrow$
& BCS$\uparrow$ & CSD$\downarrow$ & BHS$\uparrow$ & HSD$\downarrow$ & F1$\uparrow$ \\
\midrule
\multirow{8}{*}{\shortstack{General\\video-to-music\\generation}} & CMT & $32.77$ & $23.43$ & $10.51$ & $8.74$ & $14.80$ & $20.20$ & $18.43$ & $12.01$ & $11.54$ & $13.64$ \\
& Diff-Foley & $31.02$ & $23.10$ & $12.85$ & $9.51$ & $16.80$ & $23.21$ & $20.73$ & $7.86$ & $\mathbf{9.22}$ & $9.72$ \\
& M$^2$UGen & $32.10$ & $20.81$ & $43.19$ & $25.70$ & $33.24$ & $19.38$ & $11.91$ & $40.64$ & $25.83$ & $24.71$ \\
& VidMuse & $33.41$ & $18.41$ & $49.72$ & $17.41$ & $38.06$ & $19.33$ & $\mathbf{10.12}$ & $42.64$ & $15.79$ & $25.37$ \\
& GVMGen & $28.18$ & $27.77$ & $12.10$ & $16.35$ & $12.49$ & $18.54$ & $21.56$ & $10.60$ & $11.78$ & $10.82$ \\
& MTCV2M & $19.66$ & $30.66$ & $4.11$ & $8.32$ & $5.57$ & $17.88$ & $20.29$ & $16.48$ & $14.49$ & $13.73$ \\
& SONIQUE & $37.36$ & $34.44$ & $8.55$ & $12.60$ & $11.16$ & $22.69$ & $25.51$ & $19.40$ & $17.95$ & $16.16$ \\
& AudioX & $30.82$ & $25.58$ & $6.93$ & $\mathbf{6.05}$ & $10.63$ & $18.94$ & $11.47$ & $26.81$ & $16.35$ & $20.12$ \\
\midrule
\multirow{8}{*}{\shortstack{Dance-to-music\\generation}} & D2M-GAN & $34.75$ & $\mathbf{17.50}$ & $31.07$ & $10.68$ & $30.30$ & $18.66$ & $11.43$ & $17.99$ & $9.66$ & $17.28$ \\
& CDCD & $37.27$ & $24.84$ & $36.88$ & $22.09$ & $30.79$ & $22.01$ & $14.82$ & $25.42$ & $12.37$ & $18.06$ \\
& LORIS & $33.57$ & $20.38$ & $26.56$ & $14.20$ & $27.36$ & $20.36$ & $14.73$ & $26.51$ & $15.45$ & $20.97$ \\
& Dance2MusicTI & $32.99$ & $18.47$ & $39.32$ & $20.45$ & $33.79$ & $20.98$ & $11.63$ & $39.93$ & $18.10$ & $25.63$ \\
& PN-Diffusion & $34.01$ & $18.75$ & $32.44$ & $22.01$ & $29.46$ & $14.42$ & $25.55$ & $5.82$ & $10.08$ & $6.99$ \\
& UniMuMo & $30.69$ & $26.91$ & $11.16$ & $11.88$ & $13.93$ & $17.56$ & $17.79$ & $11.96$ & $13.06$ & $11.39$ \\
& AudioLDM-ControlNet & $36.39$ & $24.74$ & $15.67$ & $11.81$ & $20.11$ & $20.30$ & $12.84$ & $19.05$ & $10.67$ & $18.23$ \\
\cmidrule(l){2-12}
& \textbf{Dyna2Music (Ours)} & $\mathbf{60.44}$ & $26.57$ & $\mathbf{63.54}$ & $25.70$ & $\mathbf{55.76}$ & $\mathbf{24.06}$ & $11.44$ & $\mathbf{44.19}$ & $15.08$ & $\mathbf{29.99}$ \\
\bottomrule
\end{tabular}
}
\vspace{-7mm}
\end{table*}
\subsection{Baselines and Evaluation Protocols}
\label{subsection: baseline and evaluation}
\noindent
\textbf{Baselines.}
We organize the comparison methods into two groups.
\textit{(i) General video-to-music generation:}
CMT~\citep{di2021video}, M$^2$UGen~\citep{liu2023m2ugen}, VidMuse~\citep{tian2025vidmuse}, GVMGen~\citep{zuo2025gvmgen}, MTCV2M~\citep{wu2025controllable}, SONIQUE~\citep{zhang2025sonique}, and AudioX~\citep{tian2026audiox} represent general video soundtracking approaches.
The objective comparisons additionally include Diff-Foley~\citep{luo2023difffoley} as a video-to-audio reference.
\textit{(ii) Dance-to-music generation:}
D2M-GAN~\citep{zhu2022quantized}, CDCD~\citep{zhu2022discrete}, LORIS~\citep{yu2023long}, Dance2MusicTI~\citep{li2024dance}, PN-Diffusion~\citep{sun2025enhancing}, UniMuMo~\citep{yang2025unimumo}, and AudioLDM-ControlNet~\citep{kimura2026dance} provide comparisons with methods that condition on human motion, where UniMuMo is evaluated through its motion-to-music interface.

\begin{wraptable}{R}{0.65\textwidth}
\vspace*{0mm}
\centering
\caption{FAD (lower is better) on AIST++ and TikTok using VGGish ($v$), PANNs ($p$), and CLAP ($c$) features.
Best and second-best values are bold and underlined, respectively.}
\label{tab:current-audio-quality}
\small
\setlength{\tabcolsep}{2.5pt}
\renewcommand{\arraystretch}{1.05}
\scalebox{0.7}{
\begin{tabular}{@{}c|c|*{3}{c}|*{3}{c}@{}}
\toprule
\multirow{2}{*}{Category} & \multirow{2}{*}{Method}
& \multicolumn{3}{c|}{AIST++} & \multicolumn{3}{c}{TikTok} \\
\cmidrule(lr){3-5}\cmidrule(l){6-8}
& & FAD$_v$$\downarrow$ & FAD$_p$$\downarrow$ & FAD$_c$$\downarrow$
& FAD$_v$$\downarrow$ & FAD$_p$$\downarrow$ & FAD$_c$$\downarrow$ \\
\midrule
\multirow{8}{*}{\shortstack{General\\video-to-music\\generation}} & CMT & $10.672$ & $68.863$ & $0.958$ & $18.556$ & $80.538$ & $1.368$ \\
& Diff-Foley & $31.843$ & $184.737$ & $1.606$ & $15.147$ & $96.495$ & $1.318$ \\
& M$^2$UGen & $\underline{4.081}$ & $41.447$ & $0.808$ & $11.493$ & $60.614$ & $1.051$ \\
& VidMuse & $4.778$ & $\underline{28.645}$ & $0.975$ & $9.930$ & $52.895$ & $1.008$ \\
& GVMGen & $12.694$ & $56.269$ & $1.132$ & $10.474$ & $54.344$ & $1.147$ \\
& MTCV2M & $15.695$ & $62.076$ & $1.130$ & $9.230$ & $61.119$ & $0.952$ \\
& SONIQUE & $8.203$ & $68.007$ & $1.078$ & $11.048$ & $70.952$ & $1.116$ \\
& AudioX & $23.151$ & $96.082$ & $1.748$ & $8.539$ & $52.205$ & $0.919$ \\
\midrule
\multirow{8}{*}{\shortstack{Dance-to-music\\generation}} & D2M-GAN & $11.719$ & $71.355$ & $1.673$ & $12.885$ & $99.451$ & $1.863$ \\
& CDCD & $7.478$ & $51.817$ & $1.258$ & $6.532$ & $\underline{33.629}$ & $1.116$ \\
& LORIS & $10.733$ & $30.998$ & $1.047$ & $15.872$ & $59.243$ & $1.214$ \\
& Dance2MusicTI & $8.968$ & $49.539$ & $1.171$ & $10.146$ & $66.981$ & $1.346$ \\
& PN-Diffusion & $4.832$ & $38.041$ & $\underline{0.600}$ & $16.729$ & $71.432$ & $1.291$ \\
& UniMuMo & $10.307$ & $56.046$ & $1.157$ & $11.505$ & $66.003$ & $1.233$ \\
& AudioLDM-ControlNet & $5.632$ & $29.587$ & $0.677$ & $\underline{6.443}$ & $41.826$ & $\underline{0.771}$ \\
\cmidrule(l){2-8}
& \textbf{Dyna2Music (Ours)} & $\mathbf{2.925}$ & $\mathbf{21.259}$ & $\mathbf{0.562}$ & $\mathbf{5.342}$ & $\mathbf{32.204}$ & $\mathbf{0.493}$ \\
\bottomrule
\end{tabular}
}
\vspace{-5mm}
\end{wraptable}

\noindent
\textbf{Evaluation Protocols.}
\textit{(a) Objective Evaluation.}
We assess onset-level synchrony and audio quality 
on both datasets.
Because LORIS's one-second onset bins obscure fine timing differences, we match onset timestamps one-to-one within $\pm50$\,ms.
We report Beats Coverage Score (BCS), Beats Hit Score (BHS), and F1; definitions and across-clip variability statistics (CSD/HSD) are detailed in Appendix~\ref{sec: onset_evaluation_details}.
Audio quality is measured by Fr\'echet Audio Distance (FAD)~\citep{kilgour2018fr} using VGGish~\citep{hershey2017cnn} (FAD$_v$), PANNs~\citep{kong2020panns} (FAD$_p$), and CLAP~\citep{elizalde2023clap} (FAD$_c$) features, with lower values indicating closer agreement with reference music distributions.

\textit{(b) Subjective Evaluation.}
An AIST++ user study assesses overall musical quality (OVL) and dance-music relevance (REL).
The study protocol and results are provided in Appendix~\ref{sec: user_study}.
The supplementary material includes selected generation results from the AIST++ and TikTok test sets.

\noindent
\textbf{Implementation Details.}
{Dyna2Music operates on 64-channel music latents from a frozen waveform VAE with a temporal downsampling factor of 2048 and stereo output at $44.1$\,kHz.
The HSF-Encoder and flow-matching transformer are optimized jointly, while MotionBERT and the waveform VAE remain frozen.
We use AdamW with a learning rate of $7.5\times10^{-5}$, linear warm-up followed by linear decay, FP16 mixed precision, and gradient clipping at a norm of $1.0$.
Training is configured for up to $100$ epochs with random seed $42$, and conditioning dropout is set to $0.2$.
We set $\lambda_{\mathrm{LDC}}=1$ on both datasets.
At inference, we use exponential-moving-average weights and Euler integration over $32$ uniformly spaced flow-time points.
Architecture, sequence-length, and sampling details are provided in Appendix~\ref{sec: ldm}.}
The influence of the LDC weight on onset-level synchrony and audio quality is examined in Appendix~\ref{sec: ldc_weight_sensitivity}.

\subsection{Model Comparison}
\label{subsection: model comparison}
Tables~\ref{tab:current-rhythm-comparison} and~\ref{tab:current-audio-quality} compare onset-level synchrony and audio quality with fifteen baselines on AIST++ and TikTok.
\textbf{(i) Onset synchrony.}
Dyna2Music achieves the highest BCS, BHS, and F1 on both datasets.
Its F1 scores of $55.76$ on AIST++ and $29.99$ on TikTok exceed the strongest baselines, VidMuse and Dance2MusicTI, by $17.70$ and $4.36$ percentage points, respectively. The gains in both precision and recall indicate improved agreement with paired onsets, with a larger advantage on AIST++.
Its CSD/HSD are not the lowest, so higher mean synchrony does not imply lower variability across clips.
\textbf{(ii) Distributional audio quality.}
Dyna2Music obtains the lowest FAD in all three feature spaces on both datasets, although the strongest competing method varies by feature extractor.
This consistent ranking supports closer agreement with the reference music distributions across the three representations.

\begin{figure*}[htb]

\centering
\begin{tabular}{@{}c@{\hspace{0.03\textwidth}}c@{}}
\includegraphics[width=0.44\textwidth]{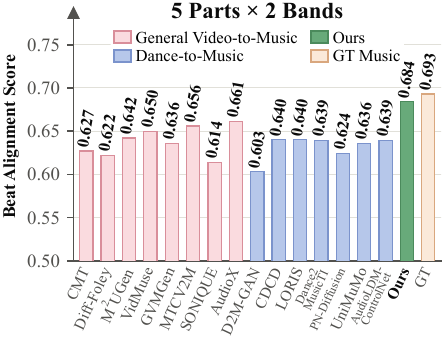} &
\includegraphics[width=0.44\textwidth]{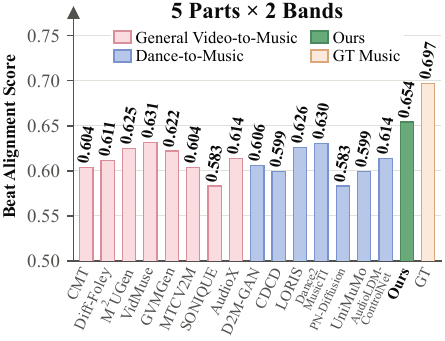} \\
{\small (a) AIST++} & {\small (b) TikTok}
\end{tabular}
\vspace*{-4mm}
\caption{Music-to-kinematic BAS ($\sigma=50$\,ms; higher is better) using the five-part, two-band motion reference on (a) AIST++ and (b) TikTok.
GT denotes paired music; colors identify method groups.}
\label{fig: bas_5p2b_comparison}
\end{figure*}

\subsection{Beat Alignment Across Motion Granularities}
\label{subsec: beat_alignment_granularities}
We evaluate music-to-kinematic BAS across the five motion-reference configurations introduced in Figure~\ref{fig: bas}.
With the five-part, two-band reference (Figure~\ref{fig: bas_5p2b_comparison}), Dyna2Music achieves the highest score among generation methods on both datasets: $0.684$ on AIST++ and $0.654$ on TikTok, exceeding the strongest baseline by $0.023$ in each case.
Both scores remain below those of the paired GT music.
It also ranks first under all four additional configurations in Appendix~\ref{sec: additional_bas_comparisons} (Figures~\ref{fig: bas_method_comparison_aist} and~\ref{fig: bas_method_comparison_tiktok}).
The consistent ranking indicates that its advantage persists across spatial partitions and frequency treatments of the evaluation reference.
These comparisons vary the scoring reference while keeping each trained model fixed.
Complementary ablations of the conditioning representation are reported in Appendix~\ref{subsec: app_conditioning_ablation}.

\subsection{Beat-Relative Rhythm Analysis}
\label{subsec: rhythm_curves}
\begingroup

We examine onset structure relative to the paired music's beat grid on $180$ AIST++ test clips spanning $10$ tracks (Figure~\ref{fig: beat_locked_curves}).
Processing details and matched-minus-mismatched comparisons are provided in Appendix~\ref{subsec: app_beat_relative} (Figure~\ref{fig: beat_locked_forest}).
\par
\endgroup

\begin{figure}[t]

\centering
\begin{tabular}{@{}c@{}}
\includegraphics[width=0.8\textwidth]{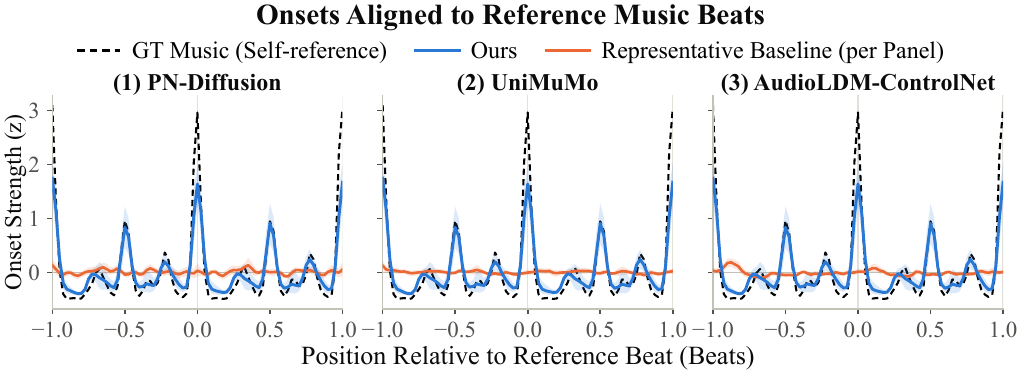} \\
{\small (a) Beat-relative onset profiles} \\[4pt]
\includegraphics[width=0.8\textwidth]{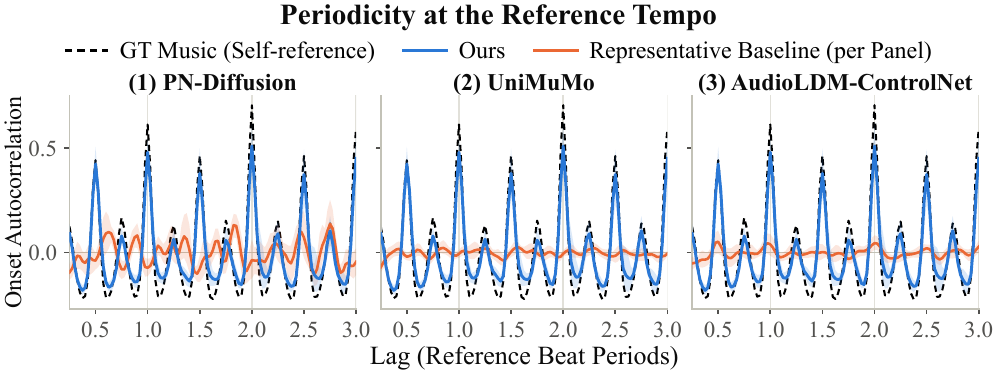} \\
{\small (b) Onset autocorrelation}
\end{tabular}
\vspace*{-2mm}
\caption{Beat-relative rhythm on AIST++: (a) standardized onset envelopes aligned to reference-music beats; (b) autocorrelation with lag normalized by the reference track's nominal beat period.
Each column compares Dyna2Music with one baseline; GT is evaluated against its own timing.
Shading denotes $95\%$ bootstrap confidence intervals over $10$ tracks.}
\label{fig: beat_locked_curves}
\end{figure}

\begingroup

Dyna2Music shows more pronounced peaks at reference beats than the representative baselines, with additional smaller peaks at half-beat subdivisions.
Its autocorrelation peaks recur at one- and two-beat lags.
These patterns support agreement with the reference rhythm in both onset placement and periodicity.
However, shifting motion conditions by approximately a quarter or half a beat leaves the onset peaks near their original phase.
This observation is consistent with a clip-onset phase prior and limits claims of local motion-dependent phase tracking.
\par
\endgroup

\subsection{Output Diversity Across Dances}
\label{subsec: diversity}
\begin{figure}[t]

\centering
\begin{tabular}{@{}c@{\hspace{0.02\textwidth}}c@{}}
\includegraphics[width=0.4\textwidth]{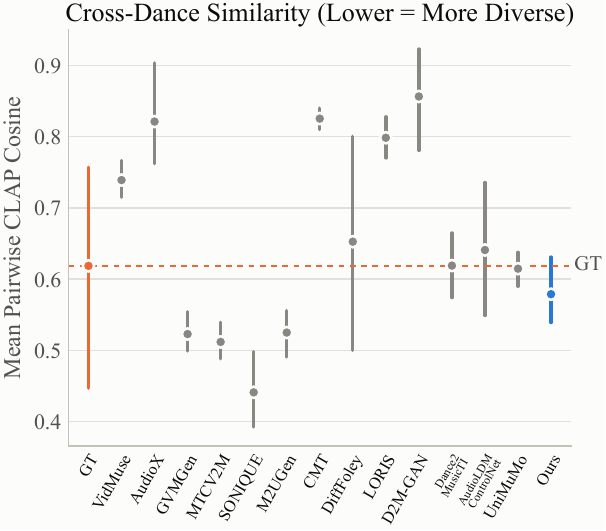} &
\includegraphics[width=0.4\textwidth]{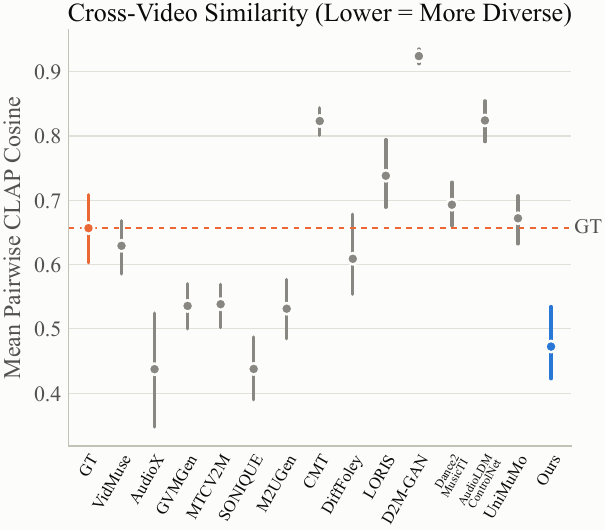} \\
{\small (a) AIST++} & {\small (b) TikTok}
\end{tabular}
\vspace*{-3mm}
\caption{Mean pairwise CLAP cosine similarity across different dances: (a) AIST++ ($180$ sequence pairs); (b) TikTok ($1{,}358$ clip pairs).
The dashed line marks the corresponding similarity of GT music.
Error bars show $95\%$ bootstrap intervals over music tracks (AIST++) or source videos (TikTok).
Higher values indicate more similar embeddings.}
\label{fig: cross_similarity}
\end{figure}
\vspace*{-3mm}
\begingroup

Figure~\ref{fig: cross_similarity} measures similarity between outputs for different dances using LAION-CLAP embeddings~\citep{wu2023large}; pair construction and estimation details are given in Appendix~\ref{subsec: app_diversity}.
Dyna2Music's mean similarity is $0.58$ on AIST++ and $0.47$ on TikTok, compared with $0.62$ and $0.66$ for GT music.
The confidence intervals overlap on AIST++, while Dyna2Music's interval lies below GT on TikTok.
In contrast, D2M-GAN and CMT have intervals entirely above GT on both benchmarks, indicating more homogeneous outputs in this feature space.
Dyna2Music therefore exhibits greater variation across dances than these baselines.

Lower similarity alone does not establish musical quality or responsiveness to motion, since outputs can differ without matching their conditioning dance.
The statistic characterizes variation across inputs; it does not measure repeated-sample diversity for a fixed dance or novelty relative to training music.
\par
\endgroup

\section{Conclusion}
\label{sec: conclusion}
We presented \textbf{Dyna2Music}, a latent flow-matching framework for variable-length dance-to-music generation.
It combines hierarchical fusion of part-specific slow-fast motion cues and pretrained joint features with latent dynamics consistency, which explicitly supervises local temporal variation in music latents.
Experiments on AIST++ and TikTok show higher onset-level synchrony and lower FAD than the evaluated baselines, with consistent BAS advantages across motion-reference configurations.
Future work will address finer motion-dependent phase control and long-range musical structure.

\clearpage
\newpage
\bibliography{iclr2024_conference}

@article{lipman2023flow,
  title={Flow matching for generative modeling},
  author={Lipman, Yaron and Chen, Ricky TQ and Ben-Hamu, Heli and Nickel, Maximilian and Le, Matt},
  journal={arXiv preprint arXiv:2210.02747},
  year={2022}
}

@article{ning2025diffrhythm,
  title={Diffrhythm: Blazingly fast and embarrassingly simple end-to-end full-length song generation with latent diffusion},
  author={Ning, Ziqian and Chen, Huakang and Jiang, Yuepeng and Hao, Chunbo and Ma, Guobin and Wang, Shuai and Yao, Jixun and Xie, Lei},
  journal={arXiv preprint arXiv:2503.01183},
  year={2025}
}

@inproceedings{li2021aichoreographer,
  title={Ai choreographer: Music conditioned 3d dance generation with aist++},
  author={Li, Ruilong and Yang, Shan and Ross, David A and Kanazawa, Angjoo},
  booktitle={2021 IEEE/CVF International Conference on Computer Vision (ICCV)},
  pages={13381--13392},
  year={2021},
  organization={Ieee}
}

@inproceedings{gan2020foley,
  title={Foley music: Learning to generate music from videos},
  author={Gan, Chuang and Huang, Deng and Chen, Peihao and Tenenbaum, Joshua B and Torralba, Antonio},
  booktitle={European Conference on Computer Vision},
  pages={758--775},
  year={2020},
  organization={Springer}
}

@inproceedings{zhu2022quantized,
  title={Quantized gan for complex music generation from dance videos},
  author={Zhu, Ye and Olszewski, Kyle and Wu, Yu and Achlioptas, Panos and Chai, Menglei and Yan, Yan and Tulyakov, Sergey},
  booktitle={European Conference on Computer Vision},
  pages={182--199},
  year={2022},
  organization={Springer}
}

@inproceedings{di2021video,
  title={Video background music generation with controllable music transformer},
  author={Di, Shangzhe and Jiang, Zeren and Liu, Si and Wang, Zhaokai and Zhu, Leyan and He, Zexin and Liu, Hongming and Yan, Shuicheng},
  booktitle={Proceedings of the 29th ACM International Conference on Multimedia},
  pages={2037--2045},
  year={2021}
}

@article{zhu2022discrete,
  title={Discrete contrastive diffusion for cross-modal music and image generation},
  author={Zhu, Ye and Wu, Yu and Olszewski, Kyle and Ren, Jian and Tulyakov, Sergey and Yan, Yan},
  journal={arXiv preprint arXiv:2206.07771},
  year={2022}
}

@inproceedings{tian2025vidmuse,
  title={Vidmuse: A simple video-to-music generation framework with long-short-term modeling},
  author={Tian, Zeyue and Liu, Zhaoyang and Yuan, Ruibin and Pan, Jiahao and Liu, Qifeng and Tan, Xu and Chen, Qifeng and Xue, Wei and Guo, Yike},
  booktitle={2025 IEEE/CVF Conference on Computer Vision and Pattern Recognition (CVPR)},
  pages={18782--18793},
  year={2025},
  organization={IEEE}
}

@inproceedings{zuo2025gvmgen,
  title={Gvmgen: A general video-to-music generation model with hierarchical attentions},
  author={Zuo, Heda and You, Weitao and Wu, Junxian and Ren, Shihong and Chen, Pei and Zhou, Mingxu and Lu, Yujia and Sun, Lingyun},
  booktitle={Proceedings of the AAAI conference on artificial intelligence},
  volume={39},
  number={21},
  pages={23099--23107},
  year={2025}
}

@inproceedings{wu2025controllable,
  title={Controllable video-to-music generation with multiple time-varying conditions},
  author={Wu, Junxian and You, Weitao and Zuo, Heda and Zhang, Dengming and Chen, Pei and Sun, Lingyun},
  booktitle={Proceedings of the 33rd ACM international conference on multimedia},
  pages={10427--10436},
  year={2025}
}

@article{liu2023m2ugen,
  title={{M$^{2}$UGen}: Multi-modal Music Understanding and Generation with the Power of Large Language Models},
  author={Liu, Shansong and Hussain, Atin Sakkeer and Wu, Qilong and Sun, Chenshuo and Shan, Ying},
  journal={arXiv preprint arXiv:2311.11255},
  year={2023}
}

@inproceedings{zhang2025sonique,
  title={Sonique: Video background music generation using unpaired audio-visual data},
  author={Zhang, Liqian and Fuentes, Magdalena},
  booktitle={ICASSP 2025-2025 IEEE International Conference on Acoustics, Speech and Signal Processing (ICASSP)},
  pages={1--5},
  year={2025},
  organization={IEEE}
}

@inproceedings{tian2026audiox,
  title={AudioX: a unified framework for anything-to-audio generation},
  author={Tian, Zeyue and Liu, Zhaoyang and Jin, Yizhu and Yuan, Ruibin and Xue, Liumeng and Tan, Xu and Chen, Qifeng and Xue, Wei and Guo, Yike},
  booktitle={International Conference on Learning Representations},
  volume={2026},
  pages={110625--110651},
  year={2026}
}

@article{luo2023difffoley,
  title={Diff-foley: Synchronized video-to-audio synthesis with latent diffusion models},
  author={Luo, Simian and Yan, Chuanhao and Hu, Chenxu and Zhao, Hang},
  journal={Advances in Neural Information Processing Systems},
  volume={36},
  pages={48855--48876},
  year={2023}
}

@inproceedings{yang2025unimumo,
  title={Unimumo: Unified text, music, and motion generation},
  author={Yang, Han and Su, Kun and Zhang, Yutong and Chen, Jiaben and Qian, Kaizhi and Liu, Gaowen and Gan, Chuang},
  booktitle={Proceedings of the AAAI Conference on Artificial Intelligence},
  volume={39},
  number={24},
  pages={25615--25623},
  year={2025}
}

@article{kimura2026dance,
  title={Dance to Music Generation leveraging Pre-training with Unpaired data and Contrastive Alignment},
  author={Kimura, Ryota and Park, Sangheon and Polouliakh, Natalia and Akama, Taketo},
  journal={arXiv preprint arXiv:2607.10537},
  year={2026}
}

@inproceedings{yu2023long,
  title={Long-term rhythmic video soundtracker},
  author={Yu, Jiashuo and Wang, Yaohui and Chen, Xinyuan and Sun, Xiao and Qiao, Yu},
  booktitle={International Conference on Machine Learning},
  pages={40339--40353},
  year={2023},
  organization={PMLR}
}

@inproceedings{sun2025enhancing,
  title={Enhancing Dance-to-Music Generation via Negative Conditioning Latent Diffusion Model},
  author={Sun, Changchang and Liu, Gaowen and Fleming, Charles and Yan, Yan},
  booktitle={Proceedings of the Computer Vision and Pattern Recognition Conference},
  pages={8321--8330},
  year={2025}
}

@article{ji2025comprehensive,
  title={A Comprehensive Survey on Generative AI for Video-to-Music Generation},
  author={Ji, Shulei and Wu, Songruoyao and Wang, Zihao and Li, Shuyu and Zhang, Kejun},
  journal={arXiv preprint arXiv:2502.12489},
  year={2025}
}

@article{han2024dance2midi,
  title={Dance2MIDI: Dance-driven multi-instrument music generation},
  author={Han, Bo and Li, Yuheng and Shen, Yixuan and Ren, Yi and Han, Feilin},
  journal={Computational Visual Media},
  volume={10},
  number={4},
  pages={791--802},
  year={2024},
  publisher={Springer}
}

@inproceedings{li2024dance,
  title={Dance-to-music generation with encoder-based textual inversion},
  author={Li, Sifei and Dong, Weiming and Zhang, Yuxin and Tang, Fan and Ma, Chongyang and Deussen, Oliver and Lee, Tong-Yee and Xu, Changsheng},
  booktitle={SIGGRAPH Asia 2024 Conference Papers},
  pages={1--11},
  year={2024}
}

@article{zhang2024dance2music,
  title={Dance2Music-Diffusion: leveraging latent diffusion models for music generation from dance videos},
  author={Zhang, Chaoyang and Hua, Yan},
  journal={EURASIP Journal on Audio, Speech, and Music Processing},
  volume={2024},
  number={1},
  pages={48},
  year={2024},
  publisher={Springer}
}

@article{aggarwal2021dance2music,
  title={Dance2music: Automatic dance-driven music generation},
  author={Aggarwal, Gunjan and Parikh, Devi},
  journal={arXiv preprint arXiv:2107.06252},
  year={2021}
}

@article{su2021does,
  title={How does it sound?},
  author={Su, Kun and Liu, Xiulong and Shlizerman, Eli},
  journal={Advances in Neural Information Processing Systems},
  volume={34},
  pages={29258--29273},
  year={2021}
}

@incollection{tan2023motion,
  title={Motion to dance music generation using latent diffusion model},
  author={Tan, Vanessa and Nam, Junghyun and Nam, Juhan and Noh, Junyong},
  booktitle={SIGGRAPH Asia 2023 Technical Communications},
  pages={1--4},
  year={2023}
}

@article{liang2024dancecomposer,
  title={DanceComposer: dance-to-music generation using a progressive conditional music generator},
  author={Liang, Xiao and Li, Wensheng and Huang, Lifeng and Gao, Chengying},
  journal={IEEE Transactions on Multimedia},
  volume={26},
  pages={10237--10250},
  year={2024},
  publisher={IEEE}
}

@inproceedings{zhu2023motionbert,
  title={Motionbert: A unified perspective on learning human motion representations},
  author={Zhu, Wentao and Ma, Xiaoxuan and Liu, Zhaoyang and Liu, Libin and Wu, Wayne and Wang, Yizhou},
  booktitle={Proceedings of the IEEE/CVF international conference on computer vision},
  pages={15085--15099},
  year={2023}
}

@inproceedings{tsuchida2019aist,
  title={AIST Dance Video Database: Multi-Genre, Multi-Dancer, and Multi-Camera Database for Dance Information Processing.},
  author={Tsuchida, Shuhei and Fukayama, Satoru and Hamasaki, Masahiro and Goto, Masataka},
  booktitle={Proceedings of the 20th International Society for Music Information Retrieval Conference},
  pages={501--510},
  year={2019}
}

@article{kilgour2018fr,
  title={Fr{\'e}chet Audio Distance: A Metric for Evaluating Music Enhancement Algorithms},
  author={Kilgour, Kevin and Zuluaga, Mauricio and Roblek, Dominik and Sharifi, Matthew},
  journal={arXiv preprint arXiv:1812.08466},
  year={2018}
}

@article{kong2020panns,
  title={Panns: Large-scale pretrained audio neural networks for audio pattern recognition},
  author={Kong, Qiuqiang and Cao, Yin and Iqbal, Turab and Wang, Yuxuan and Wang, Wenwu and Plumbley, Mark D},
  journal={IEEE/ACM Transactions on Audio, Speech, and Language Processing},
  volume={28},
  pages={2880--2894},
  year={2020},
  publisher={IEEE}
}

@inproceedings{elizalde2023clap,
  title={Clap learning audio concepts from natural language supervision},
  author={Elizalde, Benjamin and Deshmukh, Soham and Al Ismail, Mahmoud and Wang, Huaming},
  booktitle={ICASSP 2023-2023 IEEE International Conference on Acoustics, Speech and Signal Processing (ICASSP)},
  pages={1--5},
  year={2023},
  organization={IEEE}
}

@article{streijl2016mean,
  title={Mean opinion score (MOS) revisited: methods and applications, limitations and alternatives},
  author={Streijl, Robert C and Winkler, Stefan and Hands, David S},
  journal={Multimedia Systems},
  volume={22},
  number={2},
  pages={213--227},
  year={2016},
  publisher={Springer}
}

@inproceedings{siyao2022bailando,
  title={Bailando: 3d dance generation by actor-critic gpt with choreographic memory},
  author={Siyao, Li and Yu, Weijiang and Gu, Tianpei and Lin, Chunze and Wang, Quan and Qian, Chen and Loy, Chen Change and Liu, Ziwei},
  booktitle={Proceedings of the IEEE/CVF conference on computer vision and pattern recognition},
  pages={11050--11059},
  year={2022}
}

@inproceedings{raffel2014mireval,
  title={MIR\_EVAL: A Transparent Implementation of Common MIR Metrics.},
  author={Raffel, Colin and McFee, Brian and Humphrey, Eric J and Salamon, Justin and Nieto, Oriol and Liang, Dawen and Ellis, Daniel PW and Raffel, C Colin},
  booktitle={ISMIR},
  volume={10},
  pages={2014},
  year={2014}
}

@inproceedings{bock2013superflux,
  title={Maximum filter vibrato suppression for onset detection},
  author={B{\"o}ck, Sebastian and Widmer, Gerhard},
  booktitle={Proc. of the 16th Int. Conf. on Digital Audio Effects (DAFx). Maynooth, Ireland (Sept 2013)},
  volume={7},
  pages={4},
  year={2013},
  organization={Citeseer}
}

@inproceedings{mirex2009onset,
  title={Onset detection revisited},
  author={Dixon, Simon},
  booktitle={Proceedings of the 9th international conference on digital audio effects},
  volume={120},
  number={133-137},
  pages={17},
  year={2006},
  organization={Citeseer Princeton, NJ, USA}
}

@inproceedings{wu2023large,
  title={Large-scale contrastive language-audio pretraining with feature fusion and keyword-to-caption augmentation},
  author={Wu, Yusong and Chen, Ke and Zhang, Tianyu and Hui, Yuchen and Berg-Kirkpatrick, Taylor and Dubnov, Shlomo},
  booktitle={ICASSP 2023-2023 IEEE International Conference on Acoustics, Speech and Signal Processing (ICASSP)},
  pages={1--5},
  year={2023},
  organization={IEEE}
}

@inproceedings{hershey2017cnn,
  title={CNN architectures for large-scale audio classification},
  author={Hershey, Shawn and Chaudhuri, Sourish and Ellis, Daniel PW and Gemmeke, Jort F and Jansen, Aren and Moore, R Channing and Plakal, Manoj and Platt, Devin and Saurous, Rif A and Seybold, Bryan and others},
  booktitle={2017 ieee international conference on acoustics, speech and signal processing (icassp)},
  pages={131--135},
  year={2017},
  organization={IEEE}
}
\bibliographystyle{iclr2024_conference}

\clearpage
\appendix
\onecolumn
\section*{Appendix}
\setcounter{figure}{0}
\setcounter{table}{0}
\setcounter{equation}{0}
\renewcommand{\thefigure}{A\arabic{figure}}
\renewcommand{\thetable}{A\arabic{table}}
\renewcommand{\theequation}{A\arabic{equation}}
\renewcommand{\theHfigure}{appendix.\arabic{figure}}
\renewcommand{\theHtable}{appendix.\arabic{table}}
\renewcommand{\theHequation}{appendix.\arabic{equation}}

Appendix~\ref{sec: ldm} details the implementation in Section~\ref{sec: method}.
Appendices~\ref{sec: evaluation_details} and~\ref{sec: additional_bas_comparisons} supplement the evaluation protocols and BAS comparisons in Section~\ref{sec: experiment}.
The remaining appendices cover LDC-weight sensitivity, the user study, limitations, and broader impact.

\section{Implementation Details of Dyna2Music}
\label{sec: ldm}

\subsection{Motion Preprocessing and SSMD}
\label{subsec: app_motion}
This subsection details motion preprocessing for SSMD (Section~\ref{subsec: ssmd}).

\noindent\textbf{Motion inputs and joint partition.}
Input poses follow the H36M-17 convention at the native video frame rate.
Motion-energy computation and MotionBERT feature pooling use the same five-part partition, with zero-based joint indices:

\begin{center}
\begin{tabular}{ll}
\toprule
Body part & Joint indices \\
\midrule
Torso & $\{0,7,8,9,10\}$ \\
Left arm & $\{11,12,13\}$ \\
Right arm & $\{14,15,16\}$ \\
Left leg & $\{4,5,6\}$ \\
Right leg & $\{1,2,3\}$ \\
\bottomrule
\end{tabular}
\end{center}

The torso includes the root, neck, head, and other axial joints.
Velocities are consecutive pose differences multiplied by the video frame rate, with the first velocity set to zero.
The energy branch uses the supplied coordinates without per-frame torso centering.

\noindent\textbf{Filtering and temporal alignment.}
We apply fourth-order Butterworth low-pass ($1.4$\,Hz) and high-pass ($3.0$\,Hz) filters forward and backward along each velocity coordinate.
After computing mean squared speed within each part, we cap energy at the $99.5$th percentile per clip, part, and band.
Energy sequences are Fourier-resampled to the target latent length and clamped to nonnegative values.
Training uses the paired music-latent length; pose-only inference uses $\lfloor(N/f_v)(44100/2048)\rfloor$, subject to the length cap in Appendix~\ref{subsec: app_backbone}, where $f_v$ is the video frame rate.

Frozen MotionBERT extracts 512-dimensional features for all 17 joints in windows of up to 243 frames.
Features for a final segment shorter than 16 frames are computed in an overlapping window ending at the sequence boundary and cropped to that segment.
Part-wise averaging and Fourier resampling yield joint features of shape $T\times5\times512$, aligned with the slow and fast energy sequences of shape $T\times5$.
Percentile capping and nonnegative clamping apply only to energy.

\subsection{HSF-Encoder Architecture}
\label{subsec: app_hsf}
This subsection specifies the HSF-Encoder in Section~\ref{subsec: hsf}.

\noindent\textbf{Architecture.}
Layer-normalized joint features are projected to 256 dimensions; scalar energy inputs use two-layer MLPs with hidden width 128 and SiLU activation.
Energy is standardized over valid time steps per sample, part, and band, with standard deviations lower-bounded by $10^{-5}$.
Four-head attention uses band-specific query projections and shared key, value, and output projections.
Each band has a $256\rightarrow128\rightarrow1$ part-gating MLP with SiLU and a softmax over parts.
The fusion gate maps the normalized 512-dimensional branch concatenation to a scalar sigmoid output.
The final projection combines that concatenation with the 256-dimensional gated mixture, mapping 768 dimensions to 256.

\noindent\textbf{HSF embeddings and attention.}
Let $\mathbf{B}^{s}=\mathbf{S}$ and $\mathbf{B}^{f}=\mathbf{F}$, and let $\mathcal{Z}$ denote the masked temporal standardization described above.
The embeddings used in Section~\ref{subsec: hsf} are
\begin{equation}
\begin{aligned}
\widetilde{\mathbf{E}}_{n,p}
&=\mathbf{W}_{E}\mathrm{LN}(\mathbf{E}_{n,p})
 +\mathbf{e}^{\mathrm{time}}_n+\mathbf{e}^{\mathrm{part}}_p,\\
\widetilde{\mathbf{B}}^{b}_{n,p}
&=\mathrm{MLP}_{b}\!\left(\mathcal{Z}(\mathbf{B}^{b})_{n,p}\right)
 +\mathbf{e}^{\mathrm{time}}_n+\mathbf{e}^{\mathrm{part}}_p
 +\mathbf{e}^{\mathrm{band}}_b.
\end{aligned}
\label{eq: semantic}
\end{equation}
Temporal embeddings are sinusoidal; part and band embeddings are learned.
For band $b$ and head $h$, queries $\mathbf{Q}^{b}_{n,h}$ are projected from $\widetilde{\mathbf{B}}^b_n$, while keys $\mathbf{K}^{\mathrm{attn}}_{n,h}$ and values $\mathbf{V}_{n,h}$ are projected from $\widetilde{\mathbf{E}}_n$.
The operator $\mathrm{MHA}_{b}$ in Eq.~\ref{eq: slow_and_fast_2} expands as
\begin{equation}
\begin{aligned}
\mathbf{A}^{b}_{n,h}
&=\mathrm{softmax}\!\left(
\frac{\mathbf{Q}^{b}_{n,h}(\mathbf{K}^{\mathrm{attn}}_{n,h})^{\mathsf T}}{\sqrt{d_h}}
\right),\\
\mathbf{H}^{b}_n
&=\mathrm{Concat}_{h}\!\left(\mathbf{A}^{b}_{n,h}\mathbf{V}_{n,h}\right)\mathbf{W}_{O}.
\end{aligned}
\label{eq: hsf_attention_details}
\end{equation}
With four heads, $d_h=d/4$.
Each row of $\mathbf{A}^{b}_{n,h}\in\mathbb{R}^{5\times5}$ is normalized over the five key parts at the same time step.
Padded conditioning tokens are zeroed and excluded from the music generator's cross-attention.

\subsection{Music Backbone and Sequence Lengths}
\label{subsec: app_backbone}
This subsection details the generator in Section~\ref{subsec: flow_generator} and sequence lengths used in Section~\ref{subsection: baseline and evaluation}.

\noindent\textbf{Music backbone.}
The transformer has 16 blocks, hidden width 2048, 32 attention heads, and feed-forward expansion factor four.
Each block applies noncausal self-attention with rotary positional embeddings and a SiLU-gated feed-forward network, followed by separately normalized residual cross-attention to the HSF tokens.
The input concatenates 64 music-latent channels with a 512-dimensional flow-time embedding before projection to the hidden width.
The positional module uses two temporal convolutions with kernel size 31, 16 groups, and Mish activations.
Final adaptive normalization predicts scale and shift from the flow-time embedding before projection to 64 output channels.

\noindent\textbf{Waveform representation and sequence lengths.}
The frozen DiffRhythm VAE~\citep{ning2025diffrhythm} uses 64 latent channels, temporal downsampling by 2048, and stereo $44.1$\,kHz output.
Training reads precomputed reference latents; inference decodes generated latents to waveforms.
The loader requires at least 100 latent frames (approximately $4.6$\,s), excluding five of the 391 TikTok training pairs and leaving 386 eligible pairs.
All AIST++ pairs and all 53 TikTok test pairs meet this threshold.
Training sequences are padded to the configured caps: 1200 latent frames for AIST++ and 700 for TikTok (approximately $55.7$\,s and $32.5$\,s).
Valid lengths determine attention masks, energy-normalization statistics, and the averaging sets in Eqs.~\ref{eq: flow_loss} and~\ref{loss}.
Inference supports variable durations up to a configurable cap, which defaults to 1200 latent frames; longer inputs are truncated.

\subsection{Training Objective and Sampling}
\label{subsec: app_sampling}
This subsection complements Section~\ref{subsec: ldc}; optimization settings appear in Section~\ref{subsection: baseline and evaluation}.

\noindent\textbf{Flow-time sampling and velocity regression.}
For each sample $q$, we independently draw $t_q=\sigma(\xi_q)$ with $\xi_q\sim\mathcal{N}(0,1)$.
Let $\mathcal{I}$ contain the valid sample-frame pairs $(q,n)$.
The flow-matching loss averages squared velocity error over these pairs and the $d_a$ latent channels:
\begin{equation}
\mathcal{L}_{\mathrm{FM}}=
\frac{1}{d_a|\mathcal{I}|}
\sum_{(q,n)\in\mathcal{I}}
\left\|\mathbf{v}_{\theta}(\mathbf{x}_{t_q,q},t_q,\mathbf{C}_q)_n
-(\mathbf{x}_{1,q,n}-\mathbf{x}_{0,q,n})\right\|_2^2.
\label{eq: flow_loss}
\end{equation}
With probability $0.2$, a shared minibatch-level decision zeros all HSF tokens to enable classifier-free guidance.

\noindent\textbf{LDC computation.}
The single-step estimate in Eq.~\ref{eq: clean_estimate} is differentiable and recovers the paired endpoint when the predicted velocity equals $\mathbf{v}^{\star}$.
LDC reuses the flow-matching velocity prediction and is evaluated in float32 over adjacent valid frames, with the reference latent detached from the gradient graph.
Pairs involving padding are excluded.
Computation remains in latent space and requires neither beat annotations nor waveform decoding.

\noindent\textbf{Generation.}
Using exponential-moving-average weights, Euler integration maps $\mathbf{x}(0)\sim\mathcal{N}(0,\mathbf{I})$ to the latent endpoint:
\begin{equation}
\frac{\mathrm{d}\mathbf{x}(t)}{\mathrm{d}t}
=\mathbf{v}_{\theta}^{\mathrm{cfg}}(\mathbf{x}(t),t,\mathbf{C}),
\qquad
\widehat{\mathbf{y}}=\mathcal{D}(\mathbf{x}(1)),
\label{eq: sampling}
\end{equation}
where optional classifier-free guidance is implemented as
\begin{equation}
\mathbf{v}_{\theta}^{\mathrm{cfg}}
=\mathbf{v}_{\theta}(\mathbf{x},t,\mathbf{C})
+\gamma\!\left[
\mathbf{v}_{\theta}(\mathbf{x},t,\mathbf{C})
-\mathbf{v}_{\theta}(\mathbf{x},t,\mathbf{0})\right].
\label{eq: cfg}
\end{equation}
The generation scripts use $\gamma=0$, corresponding to the conditional velocity without guidance amplification.
Sampling uses 32 uniformly spaced time points $t_k=k/31$ ($k=0,\ldots,31$), giving 31 Euler updates:
\begin{equation}
\mathbf{x}^{(k+1)}=\mathbf{x}^{(k)}
+(t_{k+1}-t_k)\mathbf{v}_{\theta}^{\mathrm{cfg}}(\mathbf{x}^{(k)},t_k,\mathbf{C}).
\label{eq: euler_update}
\end{equation}
The frozen VAE decoder converts the endpoint to stereo audio; LDC and the VAE encoder are not evaluated during sampling.

\section{Evaluation Details}
\label{sec: evaluation_details}

\subsection{Onset-Level Synchrony}
\label{sec: onset_evaluation_details}
This protocol specifies the onset-level metrics in Table~\ref{tab:current-rhythm-comparison} (Section~\ref{subsection: baseline and evaluation}).

\noindent\textbf{Relation to prior evaluation.}
LORIS~\citep{yu2023long} computes Beats Coverage Score (BCS) and Beats Hit Score (BHS) from one-second bins of detected onsets.
Binning merges events and discards within-bin timing, potentially obscuring temporal mismatches.
We retain the precision/recall interpretations but match continuous onset timestamps within $\pm50$\,ms.

\noindent\textbf{Detection and matching.}
Complete audio signals are resampled to $44.1$\,kHz.
Generated and reference music use identical onset detectors: a SuperFlux-style spectral-flux envelope~\citep{bock2013superflux}, median aggregation, and a $256$-sample hop ($5.8$\,ms).
Maximum-cardinality bipartite matching in \textit{mir\_eval}~\citep{raffel2014mireval} pairs onsets one-to-one within $\pm50$\,ms~\citep{mirex2009onset}.
Scoring covers the full reference duration: excess generated events are discarded, and uncovered intervals in shorter outputs contain no generated events.

\noindent\textbf{Metrics and interpretation.}
For each clip, BCS and BHS equal the matched count divided by the generated and reference onset counts, respectively; F1 is their harmonic mean.
Table~\ref{tab:current-rhythm-comparison} reports clip-averaged BCS, BHS, and F1, and the sample standard deviations of BCS and BHS (CSD and HSD).
All five statistics are expressed as percentages.
Because chance matches depend on onset density, these scores quantify event synchrony with a particular recording and alone do not establish dance-specific rhythmic alignment.
The complementary beat-relative analysis in Section~\ref{subsec: rhythm_curves} uses the mismatched-track controls detailed below.

\subsection{Beat-Relative Onset Structure and Periodicity}
\label{subsec: app_beat_relative}
\begin{figure}[!htbp]
\centering
\includegraphics[width=0.60\textwidth]{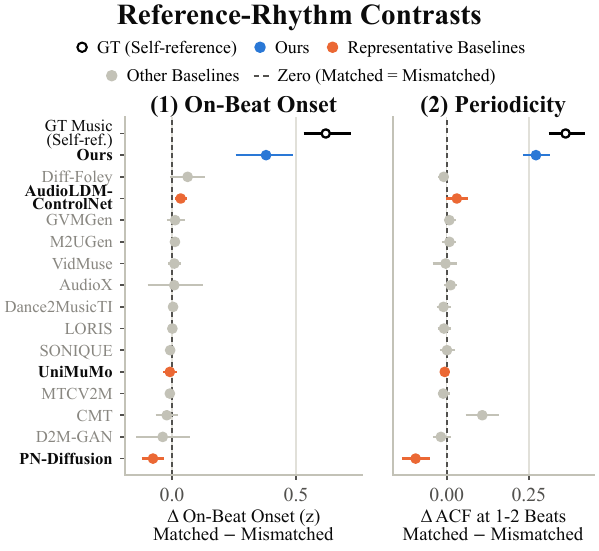}
\vspace*{-2mm}
\caption{Matched-minus-mismatched rhythm statistics on AIST++: (1) onset strength within $\pm0.1$ beats of reference beats; (2) autocorrelation within $\pm0.05$ beats of one- and two-beat lags.
Points and bars show track-averaged differences and $95\%$ bootstrap confidence intervals; zero denotes equal matched and mismatched scores.}
\label{fig: beat_locked_forest}
\end{figure}

\begingroup
\noindent\textbf{Protocol.}
Section~\ref{subsec: rhythm_curves} analyzes $234$ five-second windows from $180$ AIST++ test clips spanning $10$ music tracks.
Audio is resampled to $22.05$\,kHz; onset-strength envelopes use a $256$-sample hop and are standardized within each window.
Beat grids are estimated from complete reference recordings, with nominal tempo initializing the beat tracker.
Intervals before and after each reference beat are normalized separately to $[-1,0]$ and $[0,1]$ beats before averaging aligned envelopes (Figure~\ref{fig: beat_locked_curves}(a)).
Autocorrelation lags are normalized by the reference track's nominal beat period (Figure~\ref{fig: beat_locked_curves}(b)).
Generated audio supplies the onset envelopes; no generated beat sequence is required.

\noindent\textbf{Mismatched controls and estimation.}
Each generated window is also paired with $20$ randomly sampled beat grids from other tracks, whose nominal beat periods define the periodicity controls.
The on-beat statistic averages onset strength within $\pm0.1$ beats; the periodicity statistic averages autocorrelation within $\pm0.05$ beats of one- and two-beat lags.
Figure~\ref{fig: beat_locked_forest} reports matched-minus-mismatched differences.
We average windows within clips, clips within tracks, and then tracks with equal weight.
The $95\%$ confidence intervals use the $2.5$th and $97.5$th percentiles of $10{,}000$ track-bootstrap resamples (seed $20260921$).
GT uses its own timing as a self-reference.

\noindent\textbf{Results.}
Dyna2Music has the largest contrasts among generation methods for both on-beat onset strength ($0.379$; $95\%$ CI $[0.256,0.487]$) and periodicity ($0.272$; $[0.232,0.315]$).
AudioLDM-ControlNet shows a smaller positive on-beat contrast ($0.035$; $[0.012,0.062]$), while most baselines' intervals include zero.
CMT has positive periodicity contrast ($0.108$; $[0.058,0.160]$) but an on-beat interval containing zero, illustrating that periodicity and phase agreement provide distinct evidence.
PN-Diffusion has negative contrasts on both measures ($-0.077$ and $-0.095$).
These comparisons support reference-rhythm correspondence; local motion-dependent phase tracking remains limited, as discussed in Section~\ref{subsec: rhythm_curves}.
\par
\endgroup

\subsection{Cross-Dance Output Similarity}
\label{subsec: app_diversity}
\begingroup
The analysis in Section~\ref{subsec: diversity} and Figure~\ref{fig: cross_similarity} uses clip representations obtained by averaging normalized LAION-CLAP embeddings from five-second windows with a one-second hop.
Cosine similarity between clip representations measures output similarity.
AIST++ pairs use sequences with different music tracks and the same camera view; averaging over nine views yields $180$ sequence pairs from $20$ sequences and $10$ tracks.
For TikTok, excluding same-source pairs among $53$ clips from $43$ videos leaves $1{,}358$ pairs.

We compute the mean over eligible pairs for each method and GT music.
The $95\%$ intervals use $1{,}000$ cluster-bootstrap resamples over music tracks (AIST++) or source videos (TikTok).
Each pair receives the product of its two clusters' resampling counts, accounting for dependence from shared tracks or videos.
The statistic describes variation across inputs, rather than repeated-sample diversity for a fixed dance or novelty relative to training music.
\par
\endgroup

\section{Motion Granularity Analyses}
\label{sec: additional_bas_comparisons}
\begin{figure*}[!htbp]
\centering
\begin{tabular}{@{}c@{\hspace{0.01\textwidth}}c@{\hspace{0.01\textwidth}}c@{\hspace{0.01\textwidth}}c@{}}
\includegraphics[width=0.2425\textwidth]{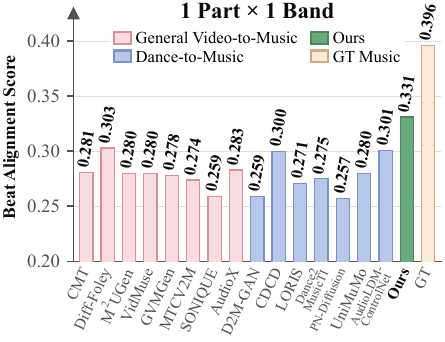} &
\includegraphics[width=0.2425\textwidth]{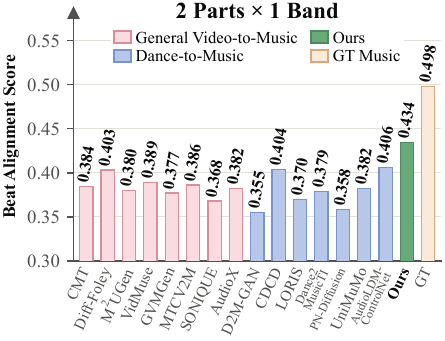} &
\includegraphics[width=0.2425\textwidth]{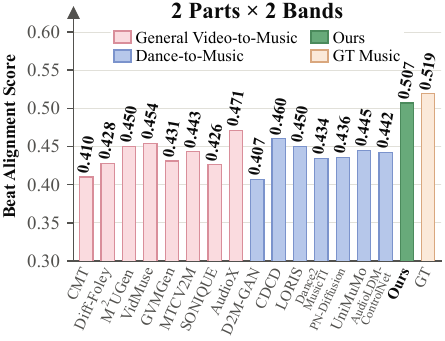} &
\includegraphics[width=0.2425\textwidth]{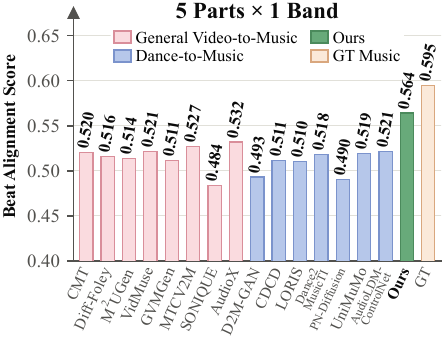} \\
{\scriptsize\begin{tabular}{@{}c@{}}(a) AIST++ \\ $1$ part $\times$ $1$ band\end{tabular}} &
{\scriptsize\begin{tabular}{@{}c@{}}(b) AIST++ \\ $2$ parts $\times$ $1$ band\end{tabular}} &
{\scriptsize\begin{tabular}{@{}c@{}}(c) AIST++ \\ $2$ parts $\times$ $2$ bands\end{tabular}} &
{\scriptsize\begin{tabular}{@{}c@{}}(d) AIST++ \\ $5$ parts $\times$ $1$ band\end{tabular}}
\end{tabular}
\vspace*{-1mm}
\caption{Music-to-kinematic BAS ($\sigma=50$\,ms) on AIST++ under four additional motion references.
Panel labels specify part and band counts; colors follow Figure~\ref{fig: bas_5p2b_comparison}.
Higher values indicate closer alignment within each configuration.}
\label{fig: bas_method_comparison_aist}
\vspace*{-3mm}
\end{figure*}

\begin{figure*}[!htbp]
\centering
\begin{tabular}{@{}c@{\hspace{0.01\textwidth}}c@{\hspace{0.01\textwidth}}c@{\hspace{0.01\textwidth}}c@{}}
\includegraphics[width=0.2425\textwidth]{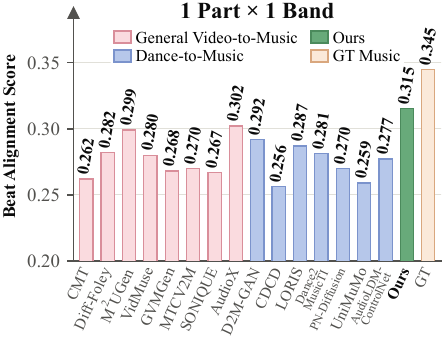} &
\includegraphics[width=0.2425\textwidth]{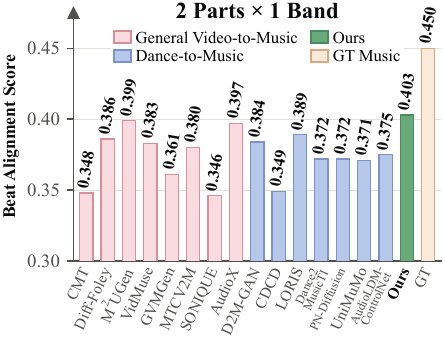} &
\includegraphics[width=0.2425\textwidth]{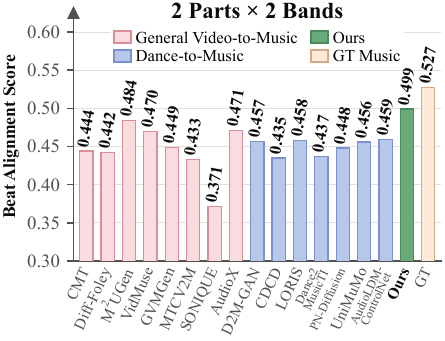} &
\includegraphics[width=0.2425\textwidth]{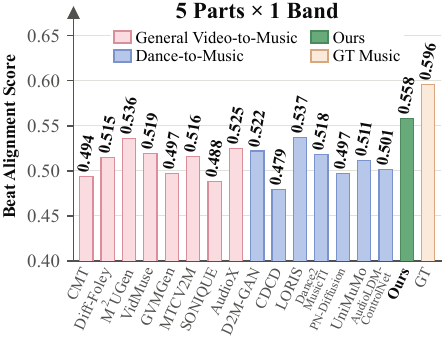} \\
{\scriptsize\begin{tabular}{@{}c@{}}(a) TikTok \\ $1$ part $\times$ $1$ band\end{tabular}} &
{\scriptsize\begin{tabular}{@{}c@{}}(b) TikTok \\ $2$ parts $\times$ $1$ band\end{tabular}} &
{\scriptsize\begin{tabular}{@{}c@{}}(c) TikTok \\ $2$ parts $\times$ $2$ bands\end{tabular}} &
{\scriptsize\begin{tabular}{@{}c@{}}(d) TikTok \\ $5$ parts $\times$ $1$ band\end{tabular}}
\end{tabular}
\vspace*{-1mm}
\caption{Music-to-kinematic BAS ($\sigma=50$\,ms) on TikTok under four additional motion references.
Panel labels specify part and band counts; colors follow Figure~\ref{fig: bas_5p2b_comparison}.
Higher values indicate closer alignment within each configuration.}
\label{fig: bas_method_comparison_tiktok}
\end{figure*}

\begingroup
Figures~\ref{fig: bas_method_comparison_aist} and~\ref{fig: bas_method_comparison_tiktok} supplement the five-part, two-band comparison in Section~\ref{subsec: beat_alignment_granularities} with four additional motion references.

\noindent\textbf{Motion references.}
One part uses the full skeleton; two parts group the torso and arms as the upper body and both legs as the lower body; five parts follow Appendix~\ref{subsec: app_motion}.
One band uses unfiltered velocity; two bands use the $1.4$\,Hz low-pass and $3.0$\,Hz high-pass components.
For each part and band, kinematic beats are local minima of mean joint speed after Gaussian smoothing ($2.5$-frame standard deviation).
Minima require prominence of $0.3$ times the curve's standard deviation and separation of at least $0.1$\,s; their union forms the motion reference.
Evaluation therefore uses speed minima as event times, whereas SSMD uses mean squared speed as conditioning energy (Eq.~\ref{eq: motion_energy}).

\noindent\textbf{Music-to-kinematic scoring.}
Music beats are detected from audio resampled to $44.1$\,kHz using a $512$-sample hop, and scoring uses the common duration of the audio and motion sequences.
Given detected music beats $\mathcal{M}$ and kinematic beats $\mathcal{K}$, the score is
\begin{equation}
\mathrm{BAS}(\mathcal{M},\mathcal{K})
=\frac{1}{|\mathcal{M}|}\sum_{m\in\mathcal{M}}
\exp\!\left(-\frac{\min_{k\in\mathcal{K}}(m-k)^2}{2\sigma^2}\right),
\qquad \sigma=0.05\,\mathrm{s}.
\label{eq: app_bas}
\end{equation}
All methods use the same reference within each configuration; GT denotes paired music scored against that reference.
Dyna2Music ranks first among generation methods in all four additional configurations on both datasets, consistent with Section~\ref{subsec: beat_alignment_granularities}.
These comparisons vary the scoring reference while keeping trained models fixed.
Raw BAS depends on reference density, so absolute scores across configurations require care in interpretation.
\par
\endgroup

\subsection{Ablation of Motion Conditioning}
\label{subsec: app_conditioning_ablation}
We compare separately trained conditioning variants using the same onset-level evaluation protocol as Table~\ref{tab:current-rhythm-comparison}.
One-, two-, and five-part variants use the whole body, upper/lower body, and the five-part partition in Appendix~\ref{subsec: app_motion}, respectively.
For the one-band variants, slow and fast energies are summed and supplied to both encoder branches; two-band variants retain separate signals.
Part merging averages both energy and joint features with weights proportional to joint counts.
Thus, these experiments alter the model's conditioning inputs, complementing the fixed-model BAS reference comparisons above.

\begin{table}[H]
\centering
\caption{Ablation of motion conditioning on AIST++ and TikTok using onset-level metrics (\%). One band denotes merged slow/fast energy. Full-model results are reproduced from Table~\ref{tab:current-rhythm-comparison}; best values within each dataset are bold.}
\label{tab: conditioning_ablation}
\small
\setlength{\tabcolsep}{6pt}
\renewcommand{\arraystretch}{1.08}
\scalebox{0.8}{
\begin{tabular*}{\linewidth}{@{\extracolsep{\fill}}lc|ccccc@{}}
\toprule
Dataset & Parts $\times$ Bands & BCS$\uparrow$ & CSD$\downarrow$ & BHS$\uparrow$ & HSD$\downarrow$ & F1$\uparrow$ \\
\midrule
\multirow{5}{*}{AIST++}
& $1\times1$ & $53.20$ & $26.31$ & $52.87$ & $25.48$ & $47.49$ \\
& $2\times1$ & $47.14$ & $24.98$ & $58.43$ & $\mathbf{22.95}$ & $47.66$ \\
& $2\times2$ & $45.92$ & $\mathbf{24.30}$ & $52.34$ & $24.31$ & $43.58$ \\
& $5\times1$ & $50.43$ & $26.10$ & $56.09$ & $26.18$ & $47.38$ \\
& $5\times2$ (full) & $\mathbf{60.44}$ & $26.57$ & $\mathbf{63.54}$ & $25.70$ & $\mathbf{55.76}$ \\
\midrule
\multirow{5}{*}{TikTok}
& $1\times1$ & $19.71$ & $\mathbf{11.29}$ & $26.66$ & $\mathbf{11.79}$ & $20.89$ \\
& $2\times1$ & $19.12$ & $11.53$ & $25.10$ & $12.45$ & $20.23$ \\
& $2\times2$ & $18.11$ & $\mathbf{11.29}$ & $26.55$ & $14.38$ & $20.01$ \\
& $5\times1$ & $21.13$ & $14.00$ & $24.77$ & $12.25$ & $20.45$ \\
& $5\times2$ (full) & $\mathbf{24.06}$ & $11.44$ & $\mathbf{44.19}$ & $15.08$ & $\mathbf{29.99}$ \\
\bottomrule
\end{tabular*}}
\end{table}

Table~\ref{tab: conditioning_ablation} shows that the full five-part, two-band model achieves the highest BCS, BHS, and F1 on both datasets.
Merging the frequency bands at five-part resolution reduces F1 by $8.38$ percentage points on AIST++ and $9.54$ on TikTok.
Retaining both bands but merging the five parts into upper/lower body reduces F1 by $12.18$ and $9.98$ points, respectively.
These results favor retaining both part and frequency distinctions in the reported models.
However, band separation at two-part resolution does not improve F1, and finer partitioning alone gives no consistent gains among the one-band variants.
The benefit therefore depends on the combined representation; differing training budgets limit attribution to conditioning structure alone.

\section{Sensitivity to the LDC Weight}
\label{sec: ldc_weight_sensitivity}
\begin{figure*}[htbp]
\centering
\begin{tabular}{@{}c@{\hspace{0.03\textwidth}}c@{}}
\includegraphics[width=0.485\textwidth]{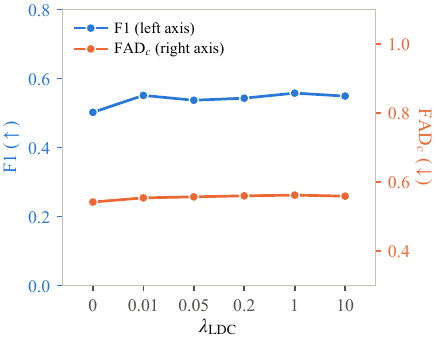} &
\includegraphics[width=0.485\textwidth]{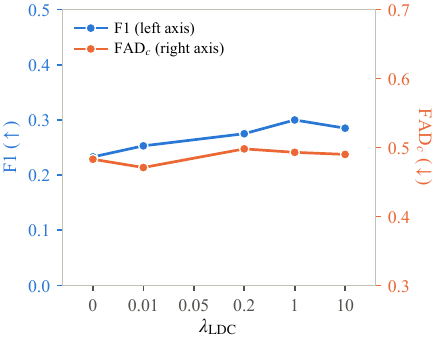} \\
{\small (a) AIST++} & {\small (b) TikTok}
\end{tabular}
\caption{Sensitivity to $\lambda_{\mathrm{LDC}}$ on (a) AIST++ and (b) TikTok.
Blue: onset-level F1 on a $0$-$1$ scale (left; higher is better).
Orange: FAD$_c$ (right; lower is better).
Horizontal positions are categorical; markers denote evaluated weights, with no $0.05$ evaluation on TikTok.}
\label{fig: ldc_weight_sensitivity}
\end{figure*}

\begingroup
Figure~\ref{fig: ldc_weight_sensitivity} examines the LDC weight in Eq.~\ref{loss} using onset-level F1 (the $\pm50$\,ms metric in Table~\ref{tab:current-rhythm-comparison}) and CLAP-based FAD$_c$.
We evaluate $\{0,0.01,0.05,0.2,1,10\}$ on AIST++ and $\{0,0.01,0.2,1,10\}$ on TikTok; zero recovers flow matching alone.

All evaluated nonzero weights improve F1 over zero on both datasets.
F1 peaks at $\lambda_{\mathrm{LDC}}=1$ ($0.5576$ on AIST++ and $0.2999$ on TikTok), compared with $0.502$ and $0.233$ at zero, and declines to $0.549$ and $0.285$ at weight $10$.
The lowest FAD$_c$ instead occurs at zero on AIST++ ($0.542$) and $0.01$ on TikTok ($0.471$), compared with $0.562$ and $0.493$ at weight $1$.
Thus, stronger LDC supervision does not monotonically improve synchrony, and the weight with the highest F1 differs from that with the lowest FAD$_c$ among the evaluated settings.
\par
\endgroup

\section{Subjective Evaluation}
\label{sec: user_study}
\begingroup
We compare Dyna2Music with AudioLDM-ControlNet and SONIQUE on AIST++, selecting these representative baselines for their generation quality (Section~\ref{subsection: baseline and evaluation}).
The study involves $10$ participants, each completing $10$ evaluation trials.
Each trial presents three versions of the same dance video, paired with soundtracks generated by the three methods.
Participants rate each version separately for overall musical quality (OVL) and dance-music relevance (REL) on a five-point scale, following~\citet{sun2025enhancing}, yielding $100$ ratings per method for each dimension.
REL assesses perceived rhythmic correspondence with the dance.
Presentation order is randomized and method identities are concealed.
We report mean opinion scores (MOS)~\citep{streijl2016mean} for each dimension, with higher values indicating better perceived quality or relevance.

Table~\ref{tab:current-subjective-comparison} reports mean OVL and REL scores of $4.3$ and $4.5$ for Dyna2Music, respectively.
These exceed the highest baseline scores by $1.0$ point for OVL (AudioLDM-ControlNet) and $1.1$ points for REL (SONIQUE).
The ratings support improvements in both perceived musical quality and dance-music relevance relative to the selected baselines.
\par
\endgroup

\begin{table}[H]
\centering
\caption{Subjective evaluation on AIST++: mean opinion scores for overall musical quality (OVL) and dance-music relevance (REL) on a $1$-$5$ scale (higher is better). Best scores are bold.}
\label{tab:current-subjective-comparison}
\small
\setlength{\tabcolsep}{8pt}
\renewcommand{\arraystretch}{1.05}
\begin{tabular*}{0.82\linewidth}{@{\extracolsep{\fill}}lcc@{}}
\toprule
Method & OVL$\uparrow$ & REL$\uparrow$ \\
\midrule
AudioLDM-ControlNet & $3.3$ & $3.0$ \\
SONIQUE & $3.1$ & $3.4$ \\
\midrule
\textbf{Dyna2Music (Ours)} & $\mathbf{4.3}$ & $\mathbf{4.5}$ \\
\bottomrule
\end{tabular*}
\end{table}

\end{document}